\def\ts     {\thinspace}
\def\msol   {\ifmmode{{\rm M}_{\odot} }\else{M$_{\odot}$}\fi}
\def\lsol   {\ifmmode{L_{\odot}}\else{$L_{\odot}$}\fi}
\def\lfir   {\ifmmode{L_{\rm FIR}}\else{$L_{\rm FIR}$}\fi}
\def\etal   {{\rm et\ts al.}}
\def\microns {\ifmmode{\mu{\rm m}}\else{$\mu$m}\fi}

\documentclass[preprint2]{aastex}
\usepackage{natbib}
\slugcomment{ApJ draft}

\shorttitle{LESS}
\shortauthors{Wei\ss, A., et al.}

\begin{document}

\title{The LABOCA Survey of the Extended Chandra Deep Field South}

\author{A. Wei\ss\altaffilmark{1}, A. Kov\'{a}cs \altaffilmark{1}, K. Coppin \altaffilmark{2}, T.R. Greve\altaffilmark{3}, F. Walter\altaffilmark{3}, Ian Smail\altaffilmark{2}, J.S. Dunlop\altaffilmark{4}, K.K. Knudsen\altaffilmark{5}, D.M. Alexander\altaffilmark{6}, F. Bertoldi\altaffilmark{5}, W.N. Brandt\altaffilmark{7}, S.C. Chapman\altaffilmark{8}, P. Cox\altaffilmark{9}, H. Dannerbauer\altaffilmark{3}, C. De Breuck\altaffilmark{10}, E. Gawiser\altaffilmark{11}, R.J. Ivison\altaffilmark{4,12}, D. Lutz\altaffilmark{13}, K.M. Menten\altaffilmark{1}, A.M. Koekemoer\altaffilmark{14}, E. Kreysa\altaffilmark{1}, P. Kurczynski\altaffilmark{11}, H.-W. Rix\altaffilmark{3}, E. Schinnerer\altaffilmark{3}, P.P. van der Werf\altaffilmark{15}}

\altaffiltext{1}{Max-Planck Institut f\"ur Radioastronomy, Auf dem H\"ugel 69, 53121 Bonn, Germany}
\altaffiltext{2}{Institute for Computational Cosmology, Durham University, South Road, Durham, DH1 3LE, UK}
\altaffiltext{3}{MPIA, K\"onigstuhl 17, 69117 Heidelberg, Germany}
\altaffiltext{4}{Institute for Astronomy, University of Edinburgh, Royal Observatory, Blackford Hill, Edinburgh, EH9 3HJ, UK }
\altaffiltext{5}{Argelander Institute for Astronomy, University of Bonn, Auf dem H\"ugel 71, D-53121 Bonn, Germany}
\altaffiltext{6}{Department of Physics, Durham University, South Road, Durham, DH1 3LE, UK}
\altaffiltext{7}{Department of Astronomy and Astrophysics, 525 Davey Lab, Pennsylvania State University, University Park, PA 16802, USA}
\altaffiltext{8}{Institute of Astronomy, Madingley Road, Cambridge, CB3 0HA, UK }
\altaffiltext{9}{Institut de Radio Astronomie Millimetrique, 300 Rue de la Piscine, Domaine Universitaire, 38406 Saint Martin d'H\'eres, France}
\altaffiltext{10}{European Southern Observatory, Karl-Schwarzschild Strasse, 85748 Garching bei M\"unchen, Germany}
\altaffiltext{11}{Yale Center for Astronomy and Astrophysics, Yale University, New Haven, CT, USA}
\altaffiltext{12}{UK Astronomy Technology Centre, Royal Observatory, Blackford Hill, Edinburgh, EH9 3HJ, UK}
\altaffiltext{13}{Max-Planck-Institut f\"ur extraterrestrische Physik, Postfach 1312, 85741 Garching, Germany}
\altaffiltext{14}{Space Telescope Science Institute, 3700 San Martin Drive, Baltimore, MD 21218, USA}
\altaffiltext{15}{Leiden Observatory, Leiden University, PO Box 9513, NL-2300 RA Leiden, the Netherlands}

\begin{abstract} We present a sensitive 870\,\micron\ survey of the
Extended Chandra Deep Field South (ECDFS) combining 310 hours of
observing time with the Large Apex BOlometer Camera (LABOCA) on the
APEX telescope. The LABOCA ECDFS Submillimetre Survey (LESS) covers
the full $30'\times30'$ field size of the ECDFS and has a uniform
noise level of $\sigma_{870\mu{\rm
m}}\approx1.2$\,mJy\,beam$^{-1}\,$. LESS is thus the largest
contiguous deep submillimetre survey undertaken to date. The noise
properties of our map show clear evidence that we are beginning to be
affected by confusion noise. We present a catalog of 126 submillimetre
galaxies (SMGs) detected with a significance level above
3.7\,$\sigma$, at which level we expect 5 false detections given our
map area of 1260 arcmin$^2$.\\ The ECDFS exhibits a deficit of bright
SMGs relative to previously studied blank fields but not of normal
star-forming galaxies that dominate the extragalactic background light
(EBL). This is in line with the underdensities observed for optically
defined high redshift source populations in the ECDFS (BzKs, DRGs,
optically bright AGN and massive K-band selected galaxies). The
differential source counts in the full field are well described by a
power law with a slope of $\alpha=-3.2$, comparable to the results
from other fields.  We show that the shape of the source counts is not
uniform across the field. Instead, it steepens in regions with low SMG
density. Towards the highest overdensities we measure a source-count
shape consistent with previous surveys. The integrated 870\micron\
flux densities of our source-count models down to $S_{870\mu{\rm
m}}=0.5$\,mJy account for $>65\%$ of the estimated EBL from {\it COBE}
measurements.  We have investigated the clustering of SMGs in the
ECDFS by means of a two-point correlation function and find evidence
for strong clustering on angular scales $<1'$ with a significance of
$3.4\,\sigma$.  Assuming a power law dependence for the correlation
function and a typical redshift distribution for the SMGs we derive a
characteristic angular clustering scale of $\theta_0=14''\pm7''$ and a
spatial correlation length of $r_0=13\pm6\,h^{-1}$\,Mpc.
\end{abstract}


\section{Introduction} One of the most significant findings of the
IRAS survey was the identification of a population of ultraluminous
infrared galaxies (ULIRGs) that emit the bulk of their bolometric
luminosity at far-IR wavelengths \citep{sanders96}.  Surveys in the
submillimetre and millimetre wavebands over the past decade have shown
that ULIRGs are much more common at high redshift compared to the
local universe \citep[e.g.][]{barger99,cowie02,borys03,webb03,greve04,laurent05,
coppin06,pope06,bertoldi07,beelen08,knudsen08,scott08,austermann09}.
These surveys show that the comoving volume density of luminous
submillimetre galaxies (SMGs) increases by a factor of $1000$ out to
$z\sim 2$ \citep{chapman05}. Therefore luminous obscured galaxies at
high redshift could dominate the total bolometric emission of galaxies
at those epochs \citep{blain99,lefloch05}

The identification and study of submillimetre galaxies has proved
challenging since their first detection.  The limited mapping speed of
typical (sub)millimetre bolometer cameras meant that only few very
bright examples have been found, although gravitational lensing
initially aided somewhat \citep[e.g.][]{smail97,ivison98}.  Attempts
to map large fields at submillimetre wavelengths have involved the use
of patchworks of small ''jiggle'' maps \citep[e.g.][]{coppin06} or
mixtures of single-bolometer photometry, small jiggle maps and shallow
scan maps used to construct a ''Super-map'' of GOODS-N
\citep{borys03,pope06}. Both of these approaches raise concerns about
the homogeneity of the resulting maps and hence the reliability of the
resulting source catalogues.  Scan maps, where the array is
continuously moved on the sky to trace out a closed pattern, should
result in much more homogeneous coverage and mapping, while at the
same time allowing for a reliable removal of the bright emission from
the atmosphere.  This technique has been used at submillimetre and
millimetre wavelengths \citep[e.g. at 350$\mu$m,][at
1100$\mu$m]{kovacs06,austermann09}, however, no deep survey have
employed such a technique in the 870-$\mu$m window where most of the
published work on SMGs has been undertaken. Drawing this distinction
between 870-$\mu$m and 1100-$\mu$m surveys may appear surprising given
the modest difference between the two wavelengths and the assumed
unstructured nature of the dust spectrum at these wavelengths.
Despite only a 25\% difference in the two wavelengths, there are hints
of significant differences in the populations identified at 870-$\mu$m
and 1100-$\mu$m \citep[e.g.][]{greve04,younger08}, although these may
in turn reflect the different mapping techniques used in individual
studies.

The advent of the new Large APEX Bolometer Camera
\citep[LABOCA,][]{siringo09}, with an instantaneous 11.4$'$ field of
view, on the 12-m APEX telescope \citep{guesten06} provided the
opportunity to undertake the first sensitive and uniform panoramic
survey of the extragalactic sky at 870$\mu$m.  To exploit this
opportunity a number of groups within the Max Planck Gesellschaft
(MPG) and the European Southern Observatory (ESO) communities proposed
a joint public legacy survey of the Extended {\it Chandra} Deep Field
South (hereafter ECDFS) to the MPG and ESO time allocation committees:
the LABOCA ECDFS submillimetre survey (LESS). The ECDFS covers a
0.5$^\circ\times$\,0.5$^\circ$ region centered on the {\it Chandra}
Deep Field South (CDFS) at RA $03^h32^m28^s.0$ Dec
$-27^{\circ}48^{'}30^{''}.0$. This field has very low far-infrared
backgrounds and good ALMA visibility and hence has become one of the
pre-eminent fields for cosmological survey science. As a result, the
ECDFS is unique in the Southern Hemisphere in the combination of area,
depth and spatial resolution of its multiwavelength coverage from
X-rays through optical, near- and mid-infrared to the far-infrared and
radio regimes. The central part of this field is coincident with the
CDFS \citep{giacconi02} which has now reached a depth of 2\,Ms
\citep{luo08} and the deep {\it Hubble Space Telescope} ({\it HST})
imaging of the GOODS-S field \citep{giovalisco04} and the {\it Hubble}
Ultra Deep Field \citep[UDF,][]{beckwith06}.  In addition to the
extremely deep observations of the central regions of this field as
part of the CDFS, GOODS and {\it Hubble} Ultra Deep Field surveys, the
full 0.5$^\circ$ field has extensive multiwavelength imaging available
including: 250-ks {\it Chandra} integrations over the whole field
\citep{lehmer05}; deep and multi-band optical imaging by COMBO-17
\citep[][2008]{wolf04} and MUSYC \citep{gawiser06} including {\it HST}
imaging for the GEMS project \citep{caldwell08}; near-infrared imaging
by MUSYC \citep{taylor09}; deep mid-infrared imaging with IRAC as part
of SIMPLE \citep{damen09} and using the MIPS instrument at 24, 70 and
$160\mu$m by FIDEL (Dickinson \etal\ in prep.).  Longer wavelength
coverage comes from BLAST \citep{devlin09} at 250, 350 and $500\mu$m
(and in the near future from {\it Herschel}), while radio coverage of
this field is reported by \citet[][]{miller08} and \citet[][]{ivison09}.

The LABOCA survey of the ECDFS adds a waveband that pin-points the
thermal emission from luminous dusty galaxies at $z\sim1$--8: a
powerful addition to this singularly well-studied region -- ideally
placed for VLT observations and early science follow-up with ALMA.
The completed LESS project provides a representative, homogeneous and
statistically-reliable sample of the SMGs with the high-quality,
multiwavelength data required to yield identifications, constrain
their redshifts, bolometric luminosities and power sources and hence
determine their contribution to the total star formation density at
high redshift. These sources can be related in unprecedented precision
to other populations of AGN and galaxies within the same volume to
understand the place of SMGs in the formation and evolution of massive
galaxies at high redshift. The survey is also sufficiently large that
it should also yield examples of rare classes of SMGs, such as very
high redshift sources, $z>4$ \citep{coppin09}. These same data also
provide submillimetre coverage of large numbers of high-redshift
galaxies and AGN to determine their bulk submillimetre properties from
the stacking analysis of sub-samples as a function of population,
redshift, environment, etc. \citep{greve09,lutz09}.  Together, these
two techniques allow us to sample two orders of magnitude in
bolometric luminosity -- from hyperluminous infrared galaxies with
$10^{13}$\,L$_\odot$ which are directly detected in the LABOCA maps,
down to luminous infrared galaxies at $10^{11}$\,L$_\odot$ which are
detected statistically through stacking.  This range in luminosity
encompasses the variety of populations expected to dominate the
bolometric emission at $z \sim 1$--3 and the cosmic submillimetre
background.

In this paper we present a detailed description of the observations,
reduction and analysis of the LABOCA observations of the ECDFS and the
resulting catalogue of submillimetre galaxies.  The observations are
described in \S2, \S3 presents our results and we discuss these in
\S4. Finally, in \S5, we give our summary and the main conclusions of
this work. We assume a cosmology with
$H_0=70$\,km\,s$^{-1}$\,Mpc$^{-1}$, $\Omega_\Lambda=0.7$ and
$\Omega_M=0.3$.

\section{Observations and data reduction \label{observations}}

Observations were carried out using the Large APEX Bolometer Camera
LABOCA \citep{siringo09} on the APEX telescope \citep{guesten06} at
Llano de Chajnantor in Chile. LABOCA is an array of 295 composite
bolometers with neutron-transmutation-doped (NTD) germanium
thermistors.  The bolometers are AC-biased and operated in total power
mode. Real-time signal processing of the 1\,kHz data stream includes
digital anti-alias filtering and down-sampling to 25\,Hz. The
radiation is coupled onto the detectors through an array of conical
feed horns whose layout leads to a double beam spaced distribution of
the individual beams in a hexagonal configuration over the $11'.4$
field of view. The center wavelength of LABOCA is 870\micron\
(345\,GHz) and its passband has a FWHM of $\sim150\micron$ (60\,GHz).
The measured angular resolution of each beam is $19''.2$ FWHM.

The observations\footnote{Programme IDs 
078.F-9028(A), 079.F-9500(A), 080.A-3023(A) and 081.F-9500(A).}
were carried out between May 2007 and November 2008
in mostly excellent weather conditions with an average precipitable
water vapor (PWV) of 0.5 mm corresponding to a zenith opacity of 0.2
at the observing wavelength. The mapping pattern was chosen to give a
uniform coverage across a $30'\times30'$ area centered at RA
$03^h32^m29^s.0$ Dec $-27^{\circ}48^{'}47^{''}.0$.  Mapping was
performed by alternating rectangular, horizontal on-the-fly (OTF)
scans with a raster of spirals pattern. OTF maps were done with a 
scanning velocity of 2\,arcmin\,s$^{-1}$ and a
spacing orthogonal to the scanning direction of $1'$. For the spiral
mode, the telescope traces in two scans spirals with radii between
$2'$ and $3'$ at 16 and 9 positions (the raster) spaced by 10$'$ in
azimuth and elevation (see Fig.\,9 in \citet{siringo09} for a plot of
this scanning pattern). The radii and spacings of the spirals were
optimized for uniform noise coverage across the $30'\times30'$ region,
while keeping telescope overheads at a minimum. The scanning speed
varies between 2 -- 3\,arcmin\,s$^{-1}$, modulating the source signals
into the useful post-detection frequency band (0.1 to 12.5 Hz) of
LABOCA, while providing at least 3 measurements per beam at the data
rate of 25 samples per second even at the highest scanning velocity.

Absolute flux calibration was achieved through observations of Mars,
Uranus and Neptune as well as secondary calibrators (V883\,Ori, NGC\,2071 and VY\,CMa) and 
was found to be accurate within 8.5\% (rms). The atmospheric attenuation was
determined via skydips every $\sim$ 2 hours as well as from
independent data from the APEX radiometer which measures the line of
sight water vapor column every minute \citep[see][for a more detailed
description]{siringo09}.  Focus settings were typically determined a
few times per night and checked during sunrise depending on the
availability of suitable sources. Pointing was checked on the nearby
quasars PMNJ0457-2324, PMNJ0106-4034 and PMNJ0403-3605 and found to be
stable within $3''$ (rms).

The data were reduced using the Bolometer array data Analysis software
(BoA, Schuller \etal\ in prep.).  Reduction steps on the time series
(time ordered data of each bolometer) include temperature drift
correction based on two "blind" bolometers (whose horns have been
sealed to block the sky signal), flat fielding, calibration, opacity
correction, flagging of unsuitable data (bad bolometers and/or data
taken outside reasonable telescope scanning velocity and acceleration
limits) as well as de-spiking. The correlated noise removal was
performed using the median signal of all bolometers in the array as
well as on groups of bolometers related by the wiring and in the
electronics \citep[see][]{siringo09}. After the de-correlation,
frequencies below 0.5\,Hz were filtered using a noise whitening
algorithm. Dead or noisy bolometers were identified based on the noise
level of the reduced time series for each detector. The number of
useful bolometers is typically $\sim 250$.  The data quality of each
scan was evaluated using the mean rms of all useful detectors before
correcting for the atmospheric attenuation (which effectively measures
the instrumental noise equivalent flux density, NEFD) and based on the
number of spikes (measuring interferences). After omitting bad data we
are left with an on-source integration time of $\sim200$\,hrs. Each
good scan was then gridded into a spatial intensity and weighting map
with a pixel size of $6''\times6''$. This pixel size ($\sim$1/3 of the
beam size) well oversamples the beam and therefore accurately
preserves the spatial information in the map. Weights are calculated
based on the rms of each time series contributing to a certain grid
point in the map. Individual maps were coadded noise-weighted. The
resulting map was used in a second iteration of the reduction to flag
those parts of the time streams with sources of a signal to noise
ratio $>3.7\sigma$. This cut off is defined by our source extraction
algorithm. The reduction with the significant sources flagged
guarantees that the source fluxes are not affected by filtering and
baseline subtraction and essentially corresponds to the very same
reduction steps that have been performed on the calibrators.

\begin{figure*}[ht] 
\centering
\includegraphics[width=12.0cm,angle=0]{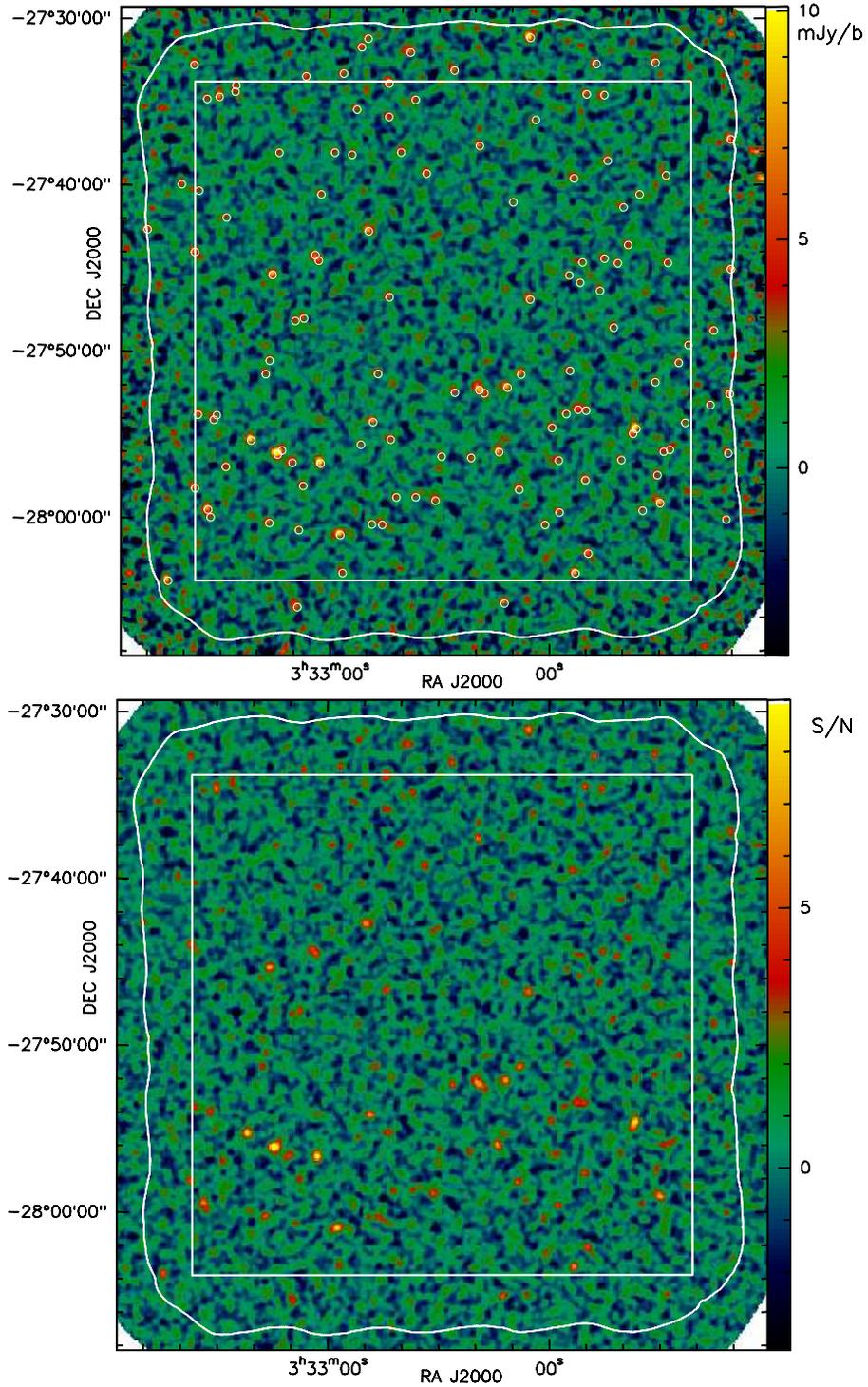} 
\caption{Flux (top) and signal to noise (bottom) map of the ECDFS at a spatial resolution of 27$''$ (beam smoothed). 
The white box shows the full $30^{'}\times 30^{'}$ of the ECDFS as defined by the GEMS project. The white 
contour shows the 1.6\,mJy\,${\rm beam}^{-1}$ noise level that has been used to define the field size for source 
extraction yielding a search area of 1260 sq. arcmins. The circles in the top panel indicate the 
location of the sources listed in 
Table\,\ref{sourcelist}.}
\label{cdfs-flux} 
\end{figure*}

To remove remaining low frequency noise artefacts we convolved the
final coadded map with a $90''$ Gaussian kernel and subtracted the
resulting large scale structures from the unsmoothed map. The
convolution kernel has been adjusted to match the low frequency excess
in the map. This step is effectively equivalent to the low frequency
behavior of an optimal point-source (Wiener) filtering operation
\citep{laurent05}. The effective decrease of the source fluxes ($\sim
5\%$) for this well defined operation has been taken into account by
scaling the fluxes accordingly.  Finally the map was beam smoothed
(convolved by the beam size of $19.2''$) to optimally filter the high
frequencies for point sources. This step reduces the spatial
resolution to $\approx27''$. The signal and signal to noise (S/N)
presentations of our final data product is shown in
Fig.\,\ref{cdfs-flux}.

To ensure that aboves reduction steps do not affect the flux
calibration of our map we performed the same reduction steps on
simulated time streams with known source fluxes and artificial
correlated and Gaussian noise.  These tests veri-
\clearpage
\noindent fied that our calibration scheme is accurate to $\sim5\%$. Furthermore we
reduced data of PSS\,2322+1944, a $z=4.1$ QSO which has been observed
during the science verification of LABOCA, in the same way as the
ECDFS data. From this measurement we find
$S_{870\microns}=21.1\pm2.5$\,mJy in good agreement with the SCUBA
measurements \citep[$22.5\pm2.5$\,mJy,][]{isaak02}.

\begin{figure}[tb]
\centering
\includegraphics[width=6.0cm,angle=0]{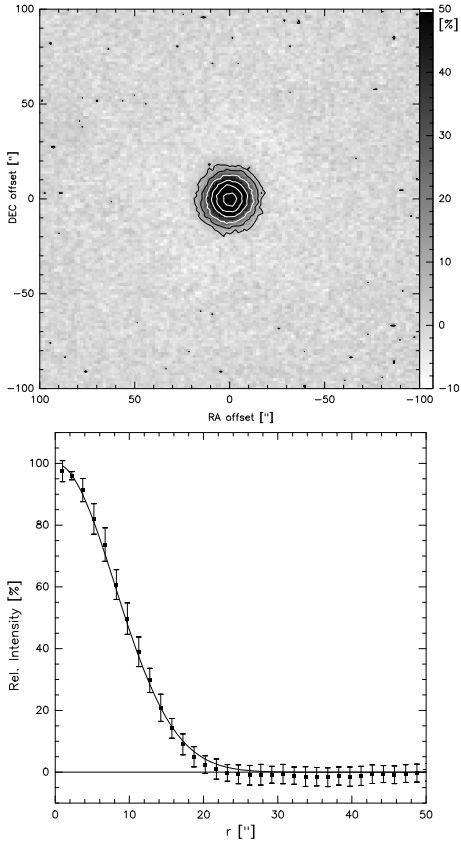}
\caption{{\it Top:} Normalized flux map of PMNJ0403-3605 (flux 590\,mJy), the most frequently used pointing 
source near the ECDFS, reduced in the same way as the ECDFS scans. Contours are shown
at 5, 15 (black) and 30, 50, 70 and 90\% (white) of the peak flux density. {\it Bottom:} Radial 
averaged beam profile. The solid line shows a Gaussian fit that yields a FWHM of $19''.2$.}
\label{beam} 
\end{figure}

Finally we determined the point spread function (PSF) in our map by
applying the same data reduction steps to PMNJ0403-3605, the most
frequently used pointing source near the ECDFS. The beam profile
(before beam smoothing) is shown in Fig.\,\ref{beam} and is well
described by a Gaussian with FWHM of $19.2''$. The faint negative
structure at radii between $25''$ and $45''$ is due to the combined
effect of the correlated noise removal, the low-frequency filtering on
the time series and the spatial large scale filtering.

\section{Results}
\subsection{Noise Properties \label{sect-noise}}

To investigate the noise properties of our LABOCA map we have created
100 pure noise realizations of the data by randomly inverting half of
the maps of individual scans during the coadding
\citep[e.g.][]{perera08,scott08}.  All image processing steps match
the real map including the large scale filtering. These so--called
''jackknife maps'' are therefore free of any astronomical signal and
at the same time represent the noise structure of the data.

\begin{figure}[t] 
\centering
\includegraphics[width=7.5cm,angle=-90]{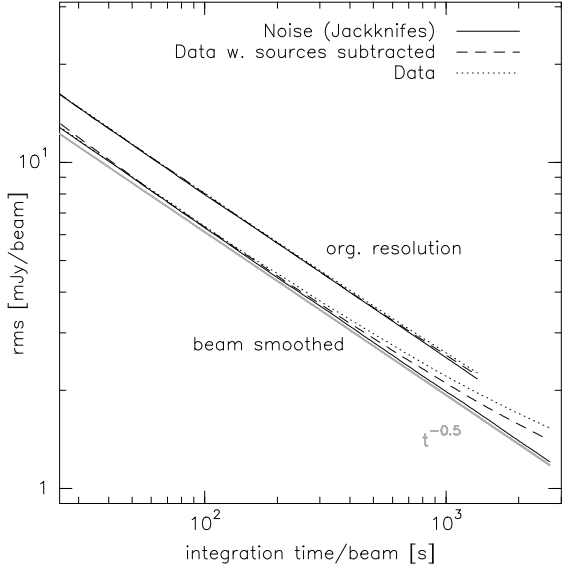}
\caption{Average noise level in the ECDFS as a function of integration time.
The upper lines represent the unsmoothed data at a spatial resolution of $19.2''$. The three lines show 
the rms noise vs. integration time for the pure noise maps (jackknifes - solid line), the flux map 
after subtraction of all sources in the catalog 
(see Sect.\,\ref{cata} - dashed line) and for the flux map including all sources (dotted line). 
The lower curves are the same but for beam-smoothed data ($27.2''$ spatial resolution).  The grey solid line has a 
slope of -0.5 and shows the expected behavior for pure noise.}
\label{cdfs-down} 
\end{figure}

We first investigate how the noise in our map integrates down with
time. This is shown in Fig.\,\ref{cdfs-down} where we plot the rms
noise level measured on the central $30^{'}\times 30^{'}$ of our map
as a function of the integration time per beam.  We have generated
down-integration curves for the pure noise realizations, the
unmodified data and for the data after subtracting the full source
catalog (see Sect.\,\ref{cata}) from each scan. All three computations
were done at the original spatial resolution as well as for the
beam-smoothed data. Ten computations with randomized scan order were
performed for each method. Fig.\,\ref{cdfs-down} shows the average of
these computations.

As expected, the pure noise realizations integrate down proportionally
to $\sqrt{1/t}$ independent of the spatial resolution.  At the
original spatial resolution of $19.2''$ the influence of the source
signals is small and even the down-integrating of the data including
all sources follows this behavior very closely. Some excess noise is
barely visible after several hundred seconds of integration time per
beam. The noise excess becomes very pronounced for the beam smoothed
data and it remains visible even after the subtraction of the source
catalog from the data.  In fact, much of the noise access (65\%)
remains in our beam-smoothed down-integrating curves after source
subtraction. This shows that the noise in our map is limited by
sources fainter than our catalog limit ($\sim4.7$\,mJy,
Sect.\,\ref{cata}) or in other words that our map starts to be
confusion limited. We have estimated the confusion noise arising from
these faint sources by fitting the down integration curve for the beam
smoothed data using
\begin{equation}
\label{down-fkt}
\sigma_{\rm obs}(t)=\sqrt{\sigma^2_{\rm n}(t)+\sigma^2_{c}}
\end{equation}
where $\sigma_{\rm obs}$, $\sigma_{\rm n}$ and $\sigma_{\rm c}$ are
the observed, instrumental/atmospheric and confusion noise terms
respectively. From this fit we derive a confusion noise of
$\sigma_{\rm c}\approx 0.9\,{\rm mJy}\,{\rm beam}^{-1}$ at $27''$
resolution. We note that this level of confusion noise is consistent
with the simulated down-integration curves based on our source counts
(see Sect.\,\ref{pd-analysis}).

\begin{figure}[t]
\centering
\includegraphics[width=6.5cm,angle=0]{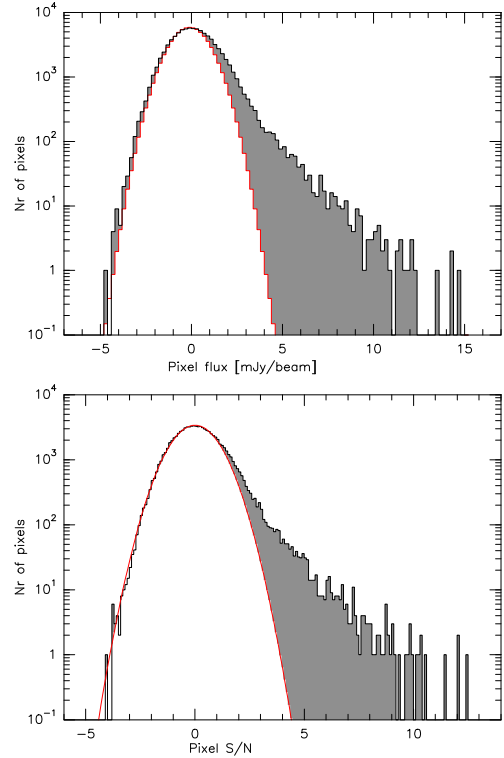}
\caption{{\it Top:} Observed beam smoothed flux histogram (black) compared to the average flux histogram of the
100 jackknifes (normalized to the same peak) in the ECDFS. The 
positive access is due to sources, the broadening of the negative part of the observed flux histogram 
is due to the confusion noise (see text). {\it Bottom:} Beam smoothed signal-to-noise (S/N)  histogram computed from the
scaled weights of the map resulting in a Gaussian with $\sigma=1$ (see Sect.\,\ref{sect-noise}).}
\label{cdfs-histo} 
\end{figure}

Given that our map rms is affected by confusion noise, the question
arises which noise level is appropriate for computing the noise
distribution across the field and the corresponding signal-to-noise
(S/N) map, which is used for source extraction. Usually, the weights
of the data (reflecting the rms weighted integration time in each
pixel) are used to derive the noise distribution. This approach,
however, is equivalent to using the jackknife noise, neglects the
confusion noise and therefore overestimates the S/N ratio. We here use
the noise based on the weights with a scaling to take the confusion
noise into account. The scaling was determined from the flux and S/N
histograms in the ECDFS. The basic principle is shown in
Fig.~\ref{cdfs-histo} (top) where we compare the average beam smoothed
flux pixel histograms of the jackknifes to a histogram of the real
map. The figure shows that the negative part of the observed histogram
is significantly broader than the jackknife histogram. This is because
inserting sources in a pure noise map (e.g. the jackknife) will not
only result in the positive flux tail, but also shift the entire flux
histogram to positive values and broaden the Gaussian part of the
distribution (because it is no longer centered around zero and sources
also fall on the formally negative part of the flux distribution).  As
the zero point (true total power information) is undetermined in our
reduction (e.g. via baseline subtraction), the observed flux histogram
is roughly centered at zero and only the broadening remains an
observable compared to the jackknife. This broadening can be used to
take the confusion noise into account. In practice, we have scaled the
rms map derived from the weights such that the negative part of the
Gaussian signal-to-noise histogram (Fig.\,\ref{cdfs-histo} bottom) has
a $\sigma$ of unity.

\begin{figure}[tb]
\centering
\includegraphics[width=6.5cm,angle=-90]{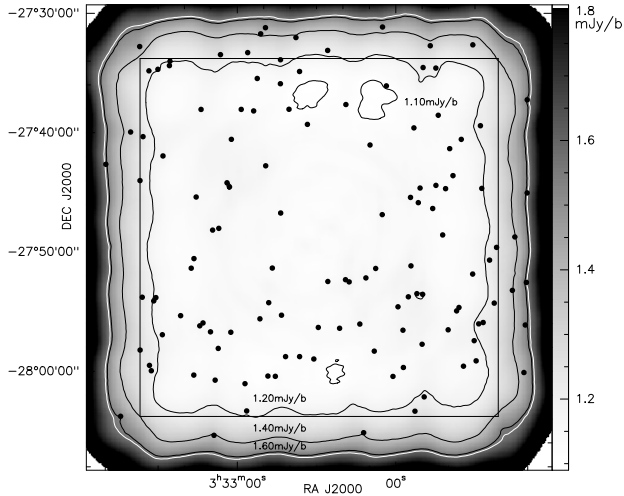}
\caption{Noise map of the ECDFS at 27$''$ resolution (beam smoothed). The circles show the
location of the 126 sources listed in Table\ \ref{sourcelist}. The black box shows
the full $30^{'}\times 30^{'}$ of the ECDFS. The white contour shows the field size that was 
used for source extraction. The black contours show the noise level 
at 1.1, 1.2, 1.4 and 1.6 mJy\,beam$^{-1}$.}
\label{rms-sources} 
\end{figure}

\begin{figure}[t] 
\centering
\includegraphics[width=7.0cm,angle=-90]{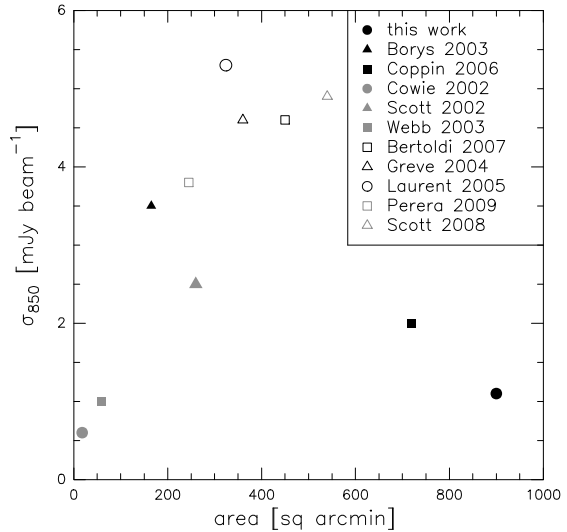}
\caption{Area and mean 850\micron\ noise level for selected mm/submm deep fields compared to LESS.
Filled and open symbols represent 850\micron\ and 1.1\,mm surveys respectively. For the later we have
scaled the noise level by a factor of 3.8, the mean flux ratio of SMGs between both wavelength \citep{greve08}.} 
\label{submm-surveys} 
\end{figure}

The resulting spatial distribution of the beam-smoothed noise level
across the imaged field is shown in Fig.\,\ref{rms-sources}. The
deepest part of our map has a noise level of 1.10 mJy\,${\rm
beam}^{-1}$, the typical rms at the edges of the $30^{'}\times 30^{'}$
field is 1.25 mJy\,${\rm beam}^{-1}$. The average noise level across
the field is 1.17 mJy\,${\rm beam}^{-1}$ with a dispersion of 40
$\mu$Jy\,${\rm beam}^{-1}$ only. For comparison with previous work we
compare the area and the noise level of selected mm/submm deep field
survey to the LESS in Fig.\,\ref{submm-surveys}.

\subsection{Source Catalog} 

\subsubsection{Source extraction algorithm}

From the beam-smoothed map we have extracted sources using the false
detection rate algorithm \citep[for a description of this method see
e.g.][]{hopkins02} of the {\sc CRUSH} package \citep{crush}. The
choice of beam smoothing becomes obvious if one considers the maximum
likelihood amplitude $A$ of a point source (i.e.~the beam $B$) fitted
at a given position ${\bf x}$ in the map $S({\bf x})$. Consider a
weighted $\chi^2$ of the fit defined in the usual way as:
\begin{equation}
\chi^2 = \sum_{\bf x'} w({\bf x'}) \left[S({\bf x'}) - A({\bf x}) B({\bf x - x'})\right]^2.
\end{equation}
Then, the $\chi^2$ minimizing condition $\partial \chi^2 / \partial A
=0$ yields a maximum-likelihood amplitude:
\begin{equation}
A({\bf x}) = \frac 
{ \sum_{\bf x'} w({\bf x'}) B({\bf x - x'})   S({\bf x'}) }
{ \sum_{\bf x'} w({\bf x'}) B^2({\bf x - x'}) },
\end{equation}
which is effectively the weighted beam smoothed image. Thus, the flux
values of the beam-smoothed map essentially measure the fitted beam
amplitudes at each map position \citep[see
also][]{serjeant03,gawiser06}.

The algorithm uses as in input parameter the allowed number of false
detections. Based on this number and the field/beam size we calculate
as an initial search criterion a detection signal-to-noise cutoff
level assuming Gaussian noise statistics. When identifying source
candidates, we allow for the possibility that the true peak may fall
between pixels by appropriately relaxing the initial search
signal-to-noise ratio. Then for each source candidate, we interpolate
the neighboring pixel values to estimate the underlying peak and its
position, and keep only candidates that meet the original detection
S/N level. We also apply the same procedure to identify negative noise
peaks. These we use to check if the false detection rate is consistent
with Gaussian noise, or to adjust the empirical noise distribution for
possible deviations thereof.

The source candidates thus identified are removed from the map using
the appropriate large-scale-structure (LSS) filtered beam
profiles. LSS filtering corrections are also calculated for each pair
of source candidates. The maps are flagged around the extracted source
positions.  After the extraction step, the zero flux level of the map
is re-estimated via the mode of the pixel distribution (which we
consider the most robust measure in this case). We also re-estimate
the width of the noise distribution using robust measures. The
extraction steps are repeated until no further candidates are
identified.

For each candidate, we calculate a detection probability based on the
number of sources detected beforehand and of the number noise peaks
found (or expected) below the corresponding inverted significance
level, under the assumption of a symmetric noise probability
distribution. Finally, the candidates are sorted in order of
decreasing detection probability. For each source we indicate the
corresponding cumulative false detection rate in
Table\,\ref{sourcelist}.

%\subsubsection{Completeness and Position accuracy \label{sect-completness}}

%To test the reliability of our extraction process and to obtain
%information on the completeness of our catalog we ran our extraction
%algorithm on 100 jackknife noise realizations after inserting
%artificial sources.
%
\begin{figure}[th] 
\centering
\includegraphics[width=6.5cm,angle=0]{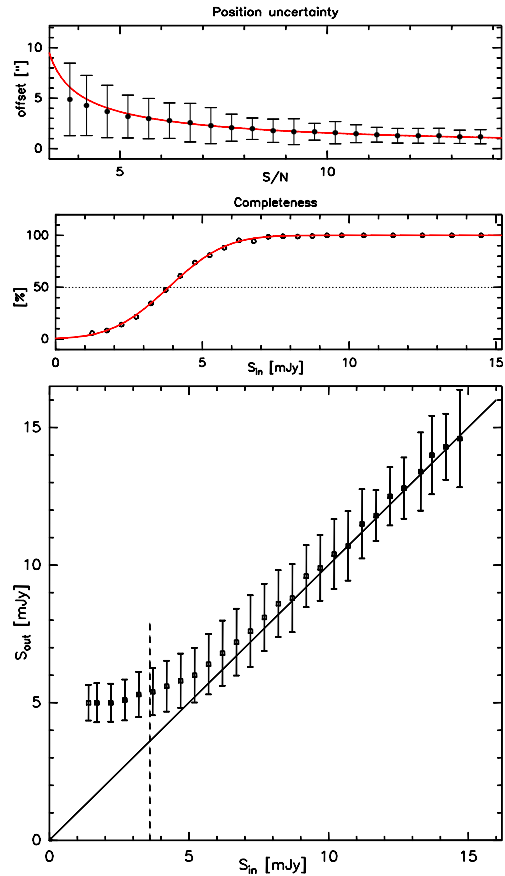}
\caption{{\it Top:} Position uncertainty as a function of observed signal-to-noise ratio. Error
bars show the rms of the extracted positions in each signal-to-noise bin, the red line shows the
expected positional uncertainties using Eq.\,B22 from \citet[][]{ivison07}.
{\it Middle and bottom:} Completeness and flux boosting as a function of intrinsic source flux 
determined from Monte-Carlo simulation. The dashed vertical line in the lower panel shows the 
lowest deboosted flux in our source catalog.}
\label{fig-completness} 
\end{figure} 

\subsubsection{Completeness and Position accuracy \label{sect-completness}}
To test the reliability of our extraction process and to obtain
information on the completeness of our catalog we ran our extraction
algorithm on 100 jackknife noise realizations after inserting
artificial sources.
The sources were added at random positions for each jackknife map and 
therefore represent an unclustered source population. For
the source flux distribution we used a power law consistent with the
differential source counts derived in
Sect.\,\ref{sect-diffcounts}. Simulated sources were inserted down to
a flux level of 1.0 mJy, i.e. well below our detection threshold for
individual source. The total number of simulated source per jackknife
map was $\sim 6000$. The large number of sources implies that the
effect of confusion noise is taken into account in our extraction. The
completeness computed from these simulations is shown in middle panel
of Fig.\,\ref{fig-completness}. The figure shows that our source
extraction is complete ($>95\%$) down to an input source flux of $\sim
6.5$\,mJy while the 50\% completeness level occurs at $\sim 4.0$\,mJy.

We also used our Monte-Carlo simulations to estimate the accuracy of
the source coordinates determined by our source extraction
algorithm. This is shown in the top panel of
Fig.~\ref{fig-completness}, where we plot the positional difference
between input and extracted position as a function of the measured
signal-to-noise ratio of the extracted sources. The mean positional
uncertainty is in good agreement with Eq.\,B22 from \citet[][red
line]{ivison07}. The scatter in the low S/N bins ($< 5$), however, is
large which implies that the positional uncertainties from aboves
equation are only correct in a statistical sense and that the true
offset of an individual source can be much larger. From the
simulations we find that for our extraction limit (S/N $>\,3.7$)
$>95\%$ of the sources will have a positional accuracy better than
$8''$.  For the brightest sources (S/N $>\,7$) the 95\% confidence
radius is $\approx3.5''$. \\ We note that because of the large number
of independent observations in our map the telescope pointing accuracy
will not affect the astrometry of our sources but result in a small
($<3\%$) spatial smearing of their signals. The accuracy of the
absolute astrometry of our map has been tested based on our stacking
analysis of a sample of near-IR selected galaxies
\citep{greve09}. These stacks give signals with up to $20\sigma$
significance centered on the expected position for different K-band
selected source populations and result in signal profiles in agreement
with the beam shape. From this we conclude that there is no evidence
for an overall mean astrometric error of our data. This is also
confirmed by a comparison of the 1.4\,GHz radio relative to the submm
positions (Ivison \etal\ in prep.).

\subsubsection{Flux boosting correction \label{deboost}}

Signal-to-noise limited source catalogs carry a selection bias from an
overabundance of sources whose apparent flux is positively enhanced by
noise \citep[e.g.][] {hogg98,scott02}.  This is shown in the bottom
panel of Fig.~\ref{fig-completness} where we compare the
extracted flux densities to the input flux densities of our
Monte-Carlo simulations.  The discrepancy between intrinsic fluxes and
their detection values becomes noticeable below $\sim 6$\,mJy.

This effect arises because the steepness of submm number counts implies 
that an observed flux $S_{\rm obs}$ more often arises from intrinsic 
fluxes $S<S_{\rm obs}$ and less often from $S>S_{\rm obs}$. The 
average flux value $\left<S\right>$ behind an observed flux $S_{\rm 
obs}$ can be calculated statistically:
\begin{equation}
\left<S\right> = \int S~ p(S | S_{\rm obs})~dS
\end{equation}
According to Bayesian theory,
$p(S_i|S_{\rm obs}) \propto p(S_{\rm obs}|S)~ p(S)$. The probability 
density $p(S_{\rm obs}|S)$ of observing a flux $S_{\rm obs}$ for an 
underlying flux $S$ is simply the noise distribution evaluated at the 
flux difference, i.e.\ $n(S-S_{\rm obs})$. For Gaussian noise $n(x)$ is 
calculated as $(\sigma \sqrt{2 \pi})^{-1}\,\exp(-x^2/2\sigma^2)$. The term 
$p(S)$ is the probability 
density of flux $S$ in the underlying noiseless flux 
distribution of the map, which accounts for the possibility of 
overlapping sources. As such, $p(S)$ is given by 
Eq.~\ref{eq:compound-dist} and is a direct product of our $P(D)$ 
analysis for the source number counts (see Sec.~\ref{pd-analysis}).

A slight complication arises because the map zeroing is biased by the 
the presence of sources below detection level. If the map zero level 
corresponds to an intrinsic flux $\delta S$, then a map flux $S$ 
really belongs to an underlying value $S+\delta S$. Fortunately, this 
$\delta S$ is also readily produced by the $P(D)$ analysis. With the 
necessary modifications in place we can calculate deboosted fluxes using
\begin{equation}
\left<S\right> = \delta S + \frac { \int  S~n(S-S_{\rm obs})~p(S+\delta 
S)~dS } { \int n(S-S_{\rm obs})~p(S+\delta S)~dS}.
\label{eq-deboost}
\end{equation}
The integration can be performed numerically. The corresponding 
uncertainty $\sigma$ of the deboosted flux can also be calculated as 
$\sigma^2 = \left<\left(S-\left<S\right>\right)^2\right>$, i.e., as

\begin{equation}
\sigma^2 = \frac { \int \left(S+\delta 
S-\left<S\right>\right)^2~n(S-S_{\rm obs})~p(S+\delta S)~dS } { \int 
n(S-S_{\rm obs})~p(S+\delta S)~dS }.
\end{equation}

The uncertainty of the deboosted flux is typically larger than the 
measurement uncertainty. The effect is more more pronounced at the lower 
fluxes.

To calculate the deboosted fluxes for each source we have used 
$p(S)$ and $\delta S$ from the single powerlaw fit from our P(D) 
analysis derived in Sect.\,\ref{pd-analysis}. Note that this is not an 
iterative process because the P(D) fitting does not require information 
on the underlying counts.

\subsubsection{The LESS source catalog \label{cata}}

While our map has a uniform noise of 1.2\,mJy beam$^{-1}$ over 900
sq. arcmin, an additional 360 sq.  arcmin has only slightly higher
noise, $<1.6$\ mJy\,beam$^{-1}$ (i.e. better than most previous
surveys, see Fig.~\ref{submm-surveys}), hence we expand our search
area to allow us to find slightly brighter sources outside the uniform
region (see Fig.~\ref{cdfs-flux}).  To construct a robust catalog we
restricted the extraction in such a way to give statistically 5 false
detections.  This yields a total of 126 sources that are listed in
Table \ref{sourcelist}.  The false detection rate (FDR) of 5 implies a
signal to noise cut above 3.7$\sigma$. The faintest sources in our
catalog have measured flux densities of 4.6\,mJy, at which level the
completeness is $>70\%$ (Fig.~\ref{fig-completness}).  In the Table we
order the individual sources by their signal-to-noise ratio and we
give their IAU name, the source position, the measured source flux
with the map noise as uncertainty, the de-boosted source flux and its
uncertainty using a Bayesian approach (see Sect.~\ref{deboost}), the
signal-to-noise ratio and the expected number of false detections for
all sources including the corresponding entry in the Table. From the
last entry in the Table it can be seen that only 10 additional sources
are included in the catalog if we increase the FDR from 3 to 5 for the
full catalog.  This implies that $~20\%$ of the additional sources are
likely to be false.  Therefore deeper source extractions would not
yield reliable information.

\subsubsection{Tests on the LESS source catalog \label{sourcetests}}

To test the reliability of our source catalog we have compared the FDR
from our Monte-Carlo simulations to the FDR expectation for our source
extraction algorithm. A source was considered to be detected if the
extraction fell within a $8''$ search radius (the maximum positional
uncertainty expected for our extraction, see \ref{sect-completness})
from the input position. The FDRs derived in both ways agree very well
with a slight tendency of our extraction algorithm to overestimate the
FDR (for 5 expected false detection we find 4 from our Monte-Carlo
simulations).

To verify that our map does not contain false sources due to artefacts
in the data or the data processing we split our observations into two
parts. This was done by splitting the randomized scan list (2370 scans
in total) into two lists with roughly equal integration time yields
two independent maps with noise levels of
$\approx1.7$\,mJy\,beam$^{-1}\,$. We then performed our source
extraction on both maps using the same significance level we used for
the generation of our source catalog (3.7$\sigma$).  This yields 59
and 60 extracted sources for map 1 and map 2, respectively which is in
excellent agreement with the expected numbers considering the
$\sqrt{2}$ increase of the noise level and the number counts derived
from the full map.

From these extractions we find that 22 sources in the LESS catalog are
detected in both submaps and that $\sim70$\% of the sources extracted
from both submaps are also in the LESS catalog. In the last entry of
Table\,\ref{sourcelist} we have indicated for each source of the LESS
catalog whether they are detected in one, both or none of the
individual maps. Note, that those sources extracted from both submaps
which are not in the LESS catalog are not necessarily false detections
as the different noise structure of the submap may boost other faint
sources above the extraction threshold.

A comparison of the sources detected on map 1 and 2 yields an overlap
of both catalogs of $\sim40\%$. For those sources that only appear in
one of the two catalogs (37/38 for map 1/2) we have extracted the SNR
peaks in a $8''$ search radius on the map where the source is not
detected and computed the probability that no source is present in
this aperture.  From this analysis we find that the probability of a
false detection exceeds 10\% for just 7(/6) source for submap 1(/2)
which is in reasonable agreement with the 5 false detections expected
from our extractions.  We furthermore analyzed the stacked signal at
the positions of the non-detections. For this we used the method
described by \citet[][]{greve09} which yields $\approx12\,\sigma$
detections for both submaps with an average flux density of
$3.5\pm0.3$\,mJy for all positions and $4.3\pm0.35$\,mJy if we exclude
the 7(/6) false detections.

Finally we have investigated the stacked
signal in maps 1 and 2 at each position at which a source is extracted
in the LESS catalog, but not in both submaps. Again the stacked
signals in both submaps give very similar results with $\approx
20\,\sigma$ detections and an average flux of $4.2\pm 0.2 $\,mJy. The
intensity of stacked signals is in good agreement with the mean
deboosted source flux of the LESS sources entering into the stack for
both maps.  These tests show, that our FDR extraction yields
reasonable results and that the average fluxes of the faint sources in
the LESS catalog can be reproduced in the stacking signals of two
independent submaps.

\begin{figure}[t]
\centering
\includegraphics[width=8.0cm,angle=-90]{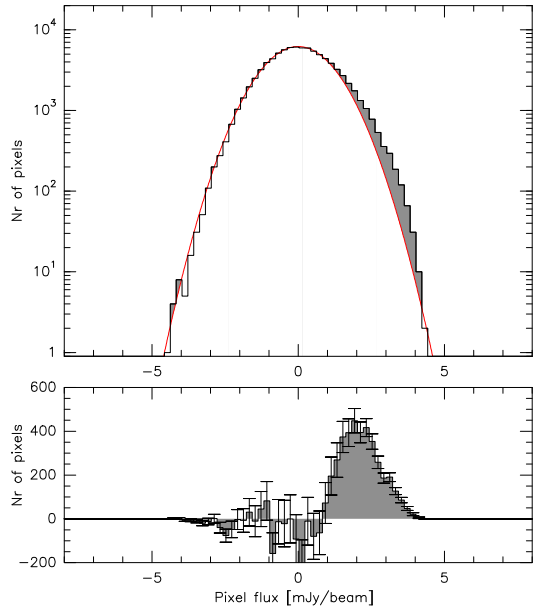}
\caption{{\it Top:} Histogram of the flux density distribution of all pixels in the central $30'\times30'$ after 
subtracting all sources from our catalog (Table \ref{sourcelist}). The curve represents a Gaussian fit to the
underlying noise histogram fitted between -5.0 and 1.0 mJy. The region shows the residual positive access.
{\it Bottom:} Linear presentation of the flux excess. Errors are 3\,$\sigma$ Poisson uncertainties.}
\label{histo-residual} 
\end{figure}

\subsection{Differential source counts \label{sect-diffcounts}}

Most studies that have addressed mm/submm number counts from
deep-field surveys have used the extracted sources, e.g. the resulting
source catalogs to fit models to the differential or integrated number
counts as a function of flux density
\citep[e.g.][]{barger99,blain99,scott02,greve04,coppin06,knudsen08,perera08,austermann09}.
This approach relies on the correct determination of the completeness
as well as on the flux de-boosting in order to extract the required
information. In particular the latter step is problematic because the
de-boosting itself requires information on the underlying source count
distribution \citep[see e.g.][]{coppin05}. Furthermore, blank field
surveys contain also information on sources fainter than the typical
cutoff levels used to extract sources \citep[e.g.][]{peacock00}.  This
is shown for our LABOCA data in Fig.\,\ref{histo-residual} where we
show the flux pixel histogram after subtracting all sources in our
source catalog from the map. The diagram shows a significant access of
pixels with positive flux densities which is due to sources below our
source extraction limit. The increasing number of noise peaks at the
same flux density level prevents an extraction directly from the map
without increasing significantly the number of false detections in the
analysis. One may, however, ignore the position information completely
and derive the underlying source count distribution directly from the
flux pixel histogram through a $P(D)$ analysis \citep[see
e.g.][]{condon74,hughes98,maloney05}.  In the bottom panel of
Fig.\,\ref{histo-residual} we show the pixel histogram of the positive
excess after subtracting a Gaussian noise distribution. The latter has
been determined by a fit to the negative part of the flux
histogram. The errors in the figure are $3\,\sigma$ Poisson
uncertainties.  From the figure it can be seen that the excess becomes
significant for flux densities above $\sim1.5$\,mJy which implies that
the source count distribution can be extracted to a much fainter level
than the limits inherent to the source extraction method.

\subsubsection{$P(D)$ analysis \label{pd-analysis}}

\begin{figure}[tb]
\centering
\includegraphics[width=8.2cm,angle=-90]{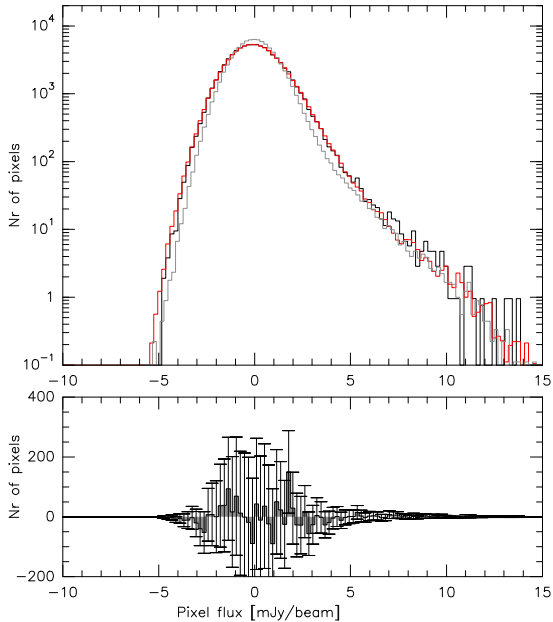}
\caption{{\it Top:} 
Logarithmic presentation of the observed flux density histogram
(black) and fit results from the $P(D)$ analysis (red) in the central
$30'\times30'$ of the E-CDFS. {\it Bottom:} Linear presentation of the
residual between the observed and fitted flux density
histogram. Errors are 3\,$\sigma$ Poisson uncertainties. The grey
histogram in the top panel shows for comparison the flux density
distribution for a Schechter function as derived for the SHADES survey
with $N'$ matched to the LESS source counts (dashed grey line in
Fig.~\ref{source-counts} left).}
\label{histo-model} 
\end{figure}

Our approach is to directly use the information from the observed flux
distribution in our map. Were there just a single source with flux $S$
present at some random location in a noiseless map, it would give rise
to a probability distribution $P(y|S)$ of pixel values $y$. If the map
contains several such randomly distributed sources with fluxes
$S_1$...$S_n$, then the probability $P(y|S_1...S_n)$ of observing a
value $y$ (in flux or S/N) can be related to the individual
probability distributions $P(y|S_i)$ arising from each individual
source and the measurement noise distribution $n(y)$ as,
\begin{equation}
%P(y|S_1...S_n) = P(y|S_1) \otimes P(y|S_2) \otimes ... \otimes P(y|S_n) \otimes n(y),
P(y|S_1...S_n) = P(y|S_1) \otimes ... \otimes P(y|S_n) \otimes n(y),
\label{eq:compound-dist}
\end{equation}
which simply follows the compounding rule for independent random
variables via convolution. As such, it can be rewritten as the product
of the complex Fourier transforms ($P \rightleftharpoons \pi$, and of
$n \rightleftharpoons \nu$) of the distributions: $\pi(S_1...S_n) =
\pi(S_1)\, \cdot \pi(S_2) ... \pi(S_n) \cdot \nu$. For example, if one
considers discrete bins $i$, each containing $N_i$ sources with flux
$S_i$, then the characteristic function $\pi$ (i.e. Fourier transform)
of the compound distribution is:
\begin{equation}
\pi = \nu \cdot \prod_i \pi(S_i)^{n_i}.
\label{eq:fourier-dist}
\end{equation}
\noindent Therefore, if one has sufficient knowledge of the underlying
noise distribution and the probability distributions due to individual
sources, then the bin counts, $n_i$, can be determined from the
observed pixel distribution with standard nonlinear
$\chi^2$-minimization techniques applied in the Fourier domain. The fitting must 
also include a parameter $\delta S$ for the true underlying flux corresponding
to map zero values. The
differential source counts are directly proportional to the bin counts
and are expressed as $dN/dS(S_i) = n_i / A \Delta_i$, in terms of the
fitted map area $A$ and the bin width $\Delta_i$.

The pixel distributions arising from sources relies on the knowledge
of the point spread function (PSF), which in our case is the closely
Gaussian shape of the LABOCA beam (Fig.~\ref{beam}), which has been
smoothed and large-scale filtered exactly like the input map. The
underlying noise distribution was obtained directly from the
jackknifed maps (Sect.\,\ref{sect-noise}), which were also smoothed
and large-scale filtered identically to the input map.

It is practical to apply the method on the signal-to-noise rather than
the flux distribution of the map, as here the noise distribution is
the narrowest when coverage is not uniform. This helps to limit the
unwanted 'smearing' of the analyzed probability distribution, thereby
improving sensitivity for determining the underlying source counts. In
our case, due to the highly uniform coverage in the central $30'
\times 30'$ area of our map used for determining source counts, the
choice between flux or S/N distribution fitting is less
critical. Nevertheless, we analysed the source counts on the
beam-smoothed S/N image.

Our choice of a non-linear fitting algorithm was based on the downhill
simplex method \citep{press86}. To use high bin resolution, necessary
given the expected steepness of the counts \citep[e.g.][]{coppin06},
and at the same time limit the number of fitted parameters to a
handful (in order to minimize covariances between them and to obtain
precise fit values for each of these) we fitted common source-counts
laws rather than the individual bin counts $n_i$.

\begin{figure*}[ht]
\centering
\includegraphics[width=14.6cm,angle=0]{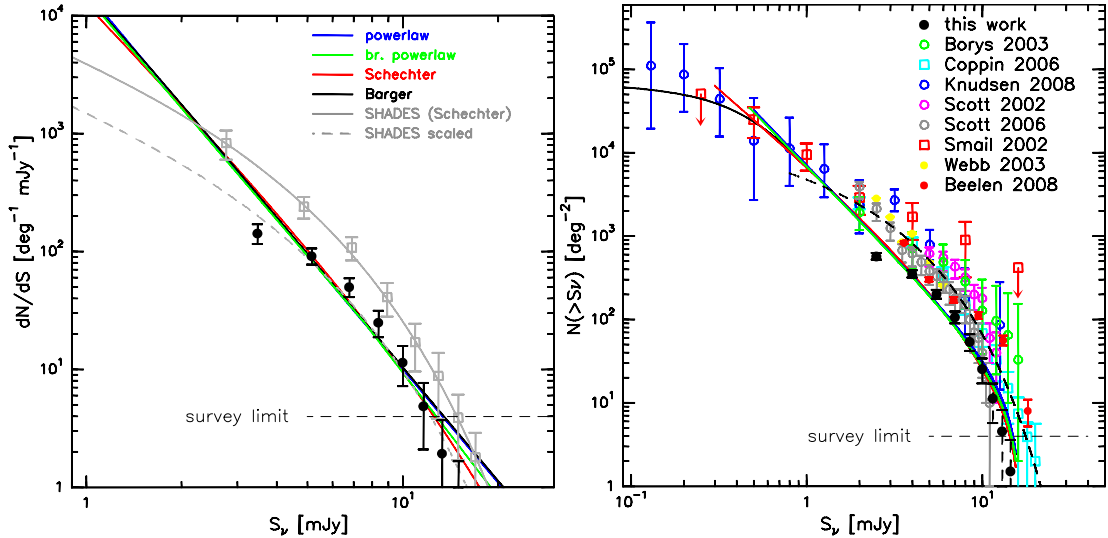}
\caption{\label{source-counts}{\it Left:} Differential source counts for the ECDFS. The colored lines show the results
from the $P(D)$ analysis with functional parameters as given in Table\,\ref{models}. The black data points show the
results from a Bayesian approach to estimate the source counts from the source catalog. The grey data points are the
differential number counts from the SHADES survey \citep{coppin06}, the grey line shows their best fitting Schechter function.
The dashed grey line is the SHADES Schechter function with $N'$ scaled to fit our Bayesian source counts. Note that this
source count function does not reproduce the observed flux density histogram of the map well (Fig.~\ref{histo-model}).
{\it Right:} Cumulative number counts for the ECDFS compared to other studies. The black data points represent the direct
sum of the differential counts shown in the left part of the figure, the solid lines are integrals over the results of the $P(D)$
analysis, the dashed line shows the best fitting Schechter function from \citet{coppin06} for SHADES.} 
\end{figure*}

We have fitted four different source count models to the observed
histogram: a single and a broken power law, a Schechter function and a
power law with constant counts at the faint end of the distribution as
suggested by \citet{barger99}. All models give comparable fits to the
observed flux histogram with reduced $\chi^2$ values about 1. The
parameters for each model are given in Table \ref{models}. As only the
Barger source count model yields finite counts for flux densities
approaching zero we also give the cutoff flux density derived from the
fitting together with the implied total extragalactic 870\micron\
background light (EBL) contribution in the Table.
Fig.\,\ref{histo-model} shows an example of the simulated flux
histogram for the best fitting Schechter function in comparison to the
observations. Within the 3$\sigma$ Poisson errors there is no
significant difference between the model and the observations. The
differential source counts are listed in Table \ref{models} and are
shown in Fig.\,\ref{source-counts}.

\subsubsection{Direct estimate of the differential source counts\label{sect-dircounts}}

To compute the differential source counts directly from the source
catalog, we have used an approach similar to that of
\citet{coppin06}. To determine the posterior flux density distribution
for each source we have used the power law function derived from our
$P(D)$ analysis (Table\,\ref{models}) as a prior for the Bayesian
deboosting of the observed flux density. The differential source
counts were calculated from the deboosted fluxes taking the Gaussian
likelihood that a source falls into a given flux bin as well as the
completeness of this bin into account. The resulting differential
source counts are shown together with the best fitting models from our
$P(D)$ analysis in Fig.~\ref{source-counts} (left).  In the figure we
show for comparison the differential source counts from the SHADES
survey \citep{coppin06} which is comparable in size, but has a noise
level $\sim2$ higher than LESS.

\begin{deluxetable}{llcccccc}
\tablecaption{Best fitting parameters of the differential source counts to the observed flux histogram.\label{models}}
\startdata
\tableline
&$^{a}$Y & S$'$ &$N'$ & $\alpha$ & $\beta$  & S$_{\rm min}$ & EBL\\
&         &[mJy] &[deg$^{-2}\,{\rm mJy}^{-1}$] & & &[mJy]&[Jy\,deg$^{-2}$]\\
\tableline
power law&$(\frac{S}{S'})^{-\alpha}$ & $^{b}$5.0 & 93&  3.2  & $-$ & 0.5& 29.1\\
br. power law&$(\frac{S}{S'})^{-\alpha}$ for $S>S'$ \\
& $(\frac{S}{S'})^{-\beta}$ for $S<S'$  & $7.6$ & 25 & 3.5 & $3.1$ &0.5 & 29.5\\
Schechter fct.&$(\frac{S}{S'})^{-\alpha}\,e^{-S/S'}$ & 10.5 &21.5 & 2.7 & $-$ & 0.3 & 33.1\\
Barger fct. &$\frac{1}{1+(\frac{S}{S'})^{\alpha}}$ & 0.56 &106000 & 3.2 & $-$ & $-$ & 32.0\\
\tableline
\enddata
\tablenotetext{a}{counts are parameterized as $\frac{dN}{dS}=N'\times Y$}
\tablenotetext{b}{$S'$ fixed to 5.0\,mJy}
\end{deluxetable}

For comparison to previous work, we derived the cumulative source
counts by directly summing over the differential source counts derived
above. The cumulative source counts are shown in comparison to other
studies in Fig.\,\ref{source-counts} (right). In this figure we also
show the integrals over the functions fitted by our $P(D)$ analysis
(Table\,\ref{models}).
%
%\begin{figure*}[tb]
%\centering
%\includegraphics[width=14.6cm,angle=0]{fig10.eps}
%\caption{\label{source-counts}{\it Left:} Differential source counts for the ECDFS. The colored lines show the results
%from the $P(D)$ analysis with functional parameters as given in Table\,\ref{models}. The black data points show the
%results from a Bayesian approach to estimate the source counts from the source catalog. The grey data points are the
%differential number counts from the SHADES survey \citep{coppin06}, the grey line shows their best fitting Schechter function.
%The dashed grey line is the SHADES Schechter function with $N'$ scaled to fit our Bayesian source counts. Note that this
%source count function does not reproduce the observed flux density histogram of the map well (Fig.~\ref{histo-model}).
%{\it Right:} Cumulative number counts for the ECDFS compared to other studies. The black data points represent the direct
%sum of the differential counts shown in the left part of the figure, the solid lines are integrals over the results of the $P(D)$
%analysis, the dashed line shows the best fitting Schechter function from \citet{coppin06} for SHADES.} 
%\end{figure*}
%
\subsection{Two point correlation}

\begin{figure}[ht]
\centering
\includegraphics[width=5.8cm,angle=-90]{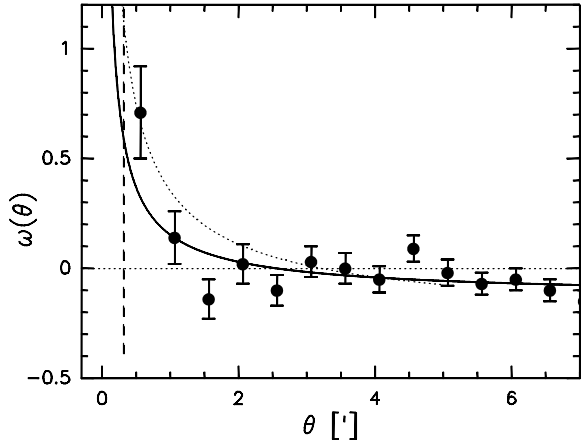}
\caption{Observed angular two point correlation function. The solid curve shows a 
power law fit to the data, the dotted line is the corresponding function derived for SMGs selected from a variety of surveys 
by \citet{scott06}.The dashed vertical line shows the size of the LABOCA beam.}
\label{angular-corr} 
\end{figure}

We have investigated the clustering properties of the SMGs in the
ECDFS by means of an angular two point correlation
function. $w(\theta)$ and its uncertainty was computed using the
\citet{landy93} estimator. The random catalog was generated from the
same simulations we used for our completeness estimate
(Sect.\,\ref{sect-completness}). To generate random positions of the
sources we used the LINUX random number generator \citep{gutterman06}.
The angular two point correlation is presented in
Fig.\,\ref{angular-corr}.  We detect positive clustering for angular
scales below $\sim1'$, although only the smallest angular scale
($20''-50''$ bin) shows statistically significant clustering
($3.4\,\sigma$).  For comparison to other studies we fit the angular
correlation by a single power law using
\begin{equation}
\label{corr-fkt}
w(\theta)=A_{w}\,(\theta^{(1-\gamma)}-C),
\end{equation}
where C accounts for the bias to lower values of the observed compared
to the true correlation \citep[see e.g.][]{brainerd98}.  As our data
are too noisy to fit all three parameters, we fixed $\gamma$ to 1.8
which has been used in many other studies
\citep[e.g][]{daddi00,farrah06,hartley08}. This yields
$A_w=0.011\pm0.0046$ and $C=12.4 \pm 2.5$ or a characteristic
clustering angle of $\theta_0=14''\pm7''$. We also calculated C
directly from our random catalog using Eq.\,22 from \citet{scott06}
and assuming $\gamma=1.8$. This yields C=4.5 and $A_w=0.007\pm0.004$
($\theta_0=7''\pm5''$) for a single parameter fit of
Eq.\,\ref{corr-fkt} to our data. These numbers demonstrate that the it
remains difficult to derive the strength of the SMG clustering from
our data but also that higher spatial resolution would greatly help to
identify close SMG pairs for a better determination of the clustering.

\section{Discussion}

\subsection{Source Counts}

The source counts derived from our $P(D)$ analysis show a reasonable
agreement with the direct estimate from the source catalog
(Fig.\,\ref{source-counts}).  We find, however, that the direct counts
yield significantly lower number counts for flux densities of $\sim
3$\,mJy and somewhat higher counts for $\sim7$\,mJy compared to the
$P(D)$ counts. The direct counts therefore suggest a deviation from a
single power law with a break between $5-7$\,mJy, similar to the break
found by \citet{coppin06} for the SHADES survey and by
\citet{knudsen08} for the Leiden SCUBA Lens Survey. This is shown in
Fig.\,\ref{source-counts} (left) where we plot the best fitting
Schechter function for the SHADES survey \citep{coppin06} with $N'$
adjusted to fit our direct counts. This comparison suggests a similar
shape (but different normalization) of the source counts between LESS
and SHADES.  A comparison of the resulting flux density histogram of
this model ($P(D)$ like diagram) to the observations, however, clearly
shows a significant deviation of the pixel counts for flux densities
below $\sim 4$\,mJy (Fig.~\ref{histo-model}).

The most likely explanation for this discrepancy is that the direct
source count method does not take multiple sources in the beam into
account, but assumes that the deboosted flux density arises from a
single source. However, from our Monte Carlo simulations we find that
this assumption is poorly justified for LESS. For differential counts
following a single power law as given in Table\,\ref{models}, we find
that ~25\% of the extracted sources in the simulations are in fact
multiple sources that fall too close together (and/or have too poor
signal to noise) to be identified as double sources. As these multiple
sources are recovered as a single source in the extraction, this
naturally leads to an underestimate of the faintest sources while
overestimating counts at higher flux densities.  This explains the
observed differences between both methods and therefore we are
confident that the $P(D)$ analysis (which does take source overlap
into account) yields more reliable results on the shape of the submm
source counts. We note that for the source counts derived from the SHADES survey 
\citet[][]{coppin06} used a correction factor determined from Monte-Carlo 
simulations to take the underestimate of the faintest sources into
account.

In any case, both methods show consistently that submm number counts
in the ECDFS are significantly lower for flux densities above
$\sim3$\,mJy compared to any other deep-fields observed at
850\,\micron\ so far. Using the SHADES number counts for comparison
(which are representative also for other fields observed with SCUBA,
see Fig.\,\ref{source-counts}) we find that the cumulative source
counts of the ECDFS are smaller by a factor of $\sim2$ for flux
densities above 3\,mJy.  This is in line with results from optical/NIR
surveys which revealed that several restframe-optical populations are
underabundant in the CDFS compared to other deep fields:
\citet{dokkum06} showed that massive K-band selected galaxies at
redshift $>2$ are underabundant by 60\%, \citet{marchesini07} reported
a mild underdensity of $z>2.5$ DRGs and \citet{dwelly06} reached
similar conclusions for optically bright AGNs at high redshifts. These
studies only covered the central $15'\times15'$ of the field where the
submm source density is even lower than in the full field (see
Sect.\,\ref{sect-varcounts} and Fig.\,\ref{fig-lss}). Similar
underdensities have also been reported in the $30'\times30'$ ECDFS for
BzKs \citep{blanc08} as well as for X-ray selected sources
\citep{lehmer05,luo08}. The central region of the ECDFS is also 
under-dense in faint ($S_{1.4{\rm GHz}} \approx 40 \mu$Jy) radio sources (E. Ibar priv. comm.).
We note, however, that the SMG underdensity is
only observed for bright sources. Due to the steep slope and the lack
of evidence for a flattening of the source counts from our P(D)
analysis for low fluxes, our number counts become consistent with
results derived for the faint end of the SMG population in
gravitationally lensed fields at flux densities of $\sim1$\,mJy
\citep{smail02,knudsen08}.

\subsection{Clustering}

Evidence for strong clustering of SMGs first emerged from
overdensities of SMGs in close proximity to other high-redshift
objects \citep[e.g.][]{ivison00,chapman01,stevens04,kneib04,beelen08}
Direct measurements of 3 and 2--D clustering of SMGs have been
presented by \citet{blain04}, \citet{greve04} and \citet{scott06}.
Due to the small number of sources typically involved in these
studies, the significance of the clustering amplitude, however,
remains marginal. Although the number of source detected by us in the
ECDFS only gives a small improvement in terms of signal to noise
compared to the previous measurements, our detection of strong
clustering in an independent submm survey greatly improves the
reliability of SMG clustering results.

Our characteristic clustering angle of $\theta_0\approx5-15''$ is
smaller than the angle of $\theta_0\approx 40-50''$ derived by
\citet{scott06} but in agreement with the predictions of merger models
\citep{vankampen05}, which predict clustering scales between $5''$
(for a hydrodynamical model) and $20''$ (for a high mass merger
model).  The difference between our results and those by
\citet{scott06} can most likely be explained by the small significance
of the clustering signal in both studies. From
Fig.\,\ref{angular-corr} it can be seen that the powerlaw fit by
\citet{scott06} is consistent with our most significant data point at
an angular distance of $40''$ and that our smaller clustering angle
mainly results from the small angular correlation for distances
between $1'-2'$ in our data. However \citet{greve04} also presented
evidence for an excess of SMG pairs with a typical separation of
$23''$ based on 1.2\,mm observations which appears to support our
findings. We note that the angular correlation function should
not be affected by the LSS filtering of our map because it refers
to the angular seperation between point sources which remains unaffected
by the filtering.

A comparison of the clustering of SMGs to other high redshift source
populations has the potential to shed more light on the evolution of
SMGs and to investigate to what kind of sources SMGs evolve once the
gas has been consumed. The clustering, however, is expected to evolve
with redshift \citep[e.g.][]{farrah06} and a conclusive comparison of
the clustering relies on a 3--D clustering analysis which requires
knowledge of the redshift distribution of our SMGs.  Assuming the same
redshift distribution for our SMGs as the distribution used by
\citet{farrah06} (a Gaussian centered at z=2.5 with a FWHM of 1.2,
similar to the redshift distribution of SMGs derived by
\citet[][]{chapman05}) our data suggests a correlation length of
$r_0=10\pm6\,h^{-1}$\,Mpc and $r_0=13\pm6\,h^{-1}$\,Mpc for our single
and two parameter fit, respectively. These values are larger than, but
consistent within the errors with, the correlation length of
$r_0=6.9\pm2.1\,h^{-1}$\,Mpc derived by \citet{blain04} for SMGs and
in good agreement with the correlation length of
$r_0=14.4\pm2.\,h^{-1}$\,Mpc derived for 24\,\micron\ selected ULIRGs
at $z=2-3$ \citep{farrah06}.

Most models for the evolution of overdensities over time predict an
increasing correlation length for decreasing redshift \citep[for a
collection of evolution models see e.g.][]{overzier03}. This suggests
that the successors of SMGs could be associated with clusters of
galaxies ($r_0\approx 20\,h^{-1}$\,Mpc) at the present
epoch. Comparing our clustering strength to the correlation length of
dark matter (DM) halos as a function of redshift
\citep[e.g.][]{matarrese97} furthermore suggests that SMGs reside in
more massive ($\sim10^{13-14}\,\msol$) DM halos than other high-z
source populations such as LBGs and QSOs. Given the uncertainties of
our measurement, the unknown redshift distribution of our SMGs and the
model dependence of the DM clustering these numbers are quite
uncertain but support the conclusions of previous studies
\citep{blain04,farrah06}.

\subsection{Spatial variations of the source counts \label{sect-varcounts}}

Motivated by the strong clustering detected in the distribution of the
\begin{figure}[ht]
\centering
\includegraphics[width=6.5cm,angle=-90]{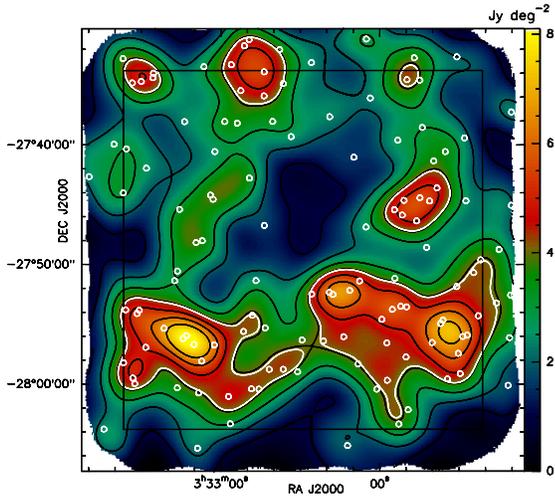}
\caption{870 \microns\ flux density distribution smoothed to 5$'$ spatial resolution. The flux 
density is show in Jy\,deg$^{-2}$. The black square shows the central $30'\times30'$ 
of the ECDFS. The white circles mark the location of the 126 sources listed in Table\ \ref{sourcelist}.
Black contours are shown from 1 to 7\,Jy \,deg$^{-2}$ spaced by 1\,Jy \,deg$^{-2}$. The white 
contour is the 4.1 Jy\,deg$^{-2}$ level that has been used to define the SMG over and underdensity region.}
\label{fig-lss}
\end{figure}
submm sources we also investigated the spatial variations of the
source counts across the map. To distinguish between regions with
potential over and underdensities of submm sources we have used the
integrated extragalactic 870\micron\ flux detected in the map as
reference. The variations were calculated at spatial resolution of
$5'$ by convolving the flux map by a Gaussian kernel. The resulting
flux distribution is shown in Fig.\,\ref{fig-lss} in units of
Jy\,deg$^{-2}$ and reflects mainly the large scale distribution of the
individually detected sources as faint extended emission has been
removed by our optimal point-source filtering operation (see
Sect.\,\ref{observations}). We have used the 4.1 Jy\,deg$^{-2}$
contour to divide our source catalog into two sub-catalogs which we
call for convenience the 'sparse' and the 'dense' region in the
following. The contour was chosen to give roughly an equal number of
detected sources for both catalogs. The source selection based on the
large scale flux distribution emphasizes the difference of the source
counts between over and underdensities and is no longer representative
for the counts in a blank field survey. 

\begin{figure}[th]
\centering
\includegraphics[width=6.1cm,angle=0]{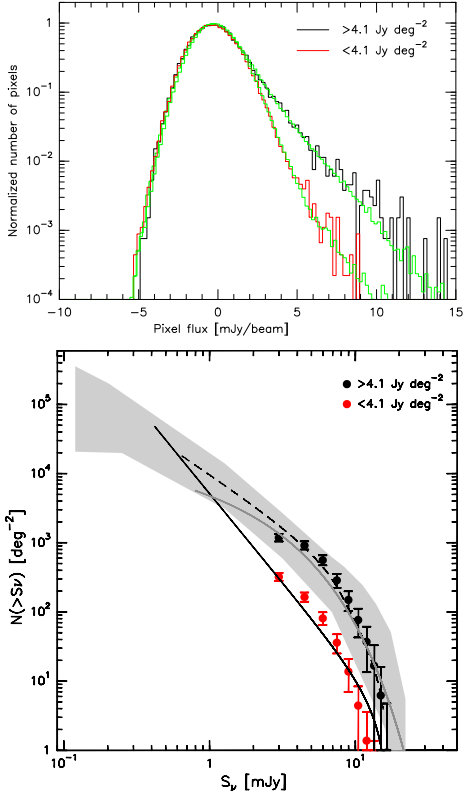}
\caption{Variation of the source counts with source surface density: {\it Top:} 
The black and the red histograms show the pixel flux distribution for
the dense and sparse region (Fig.\,\ref{fig-lss}).  The green curves
show the results from our $P(D)$ analysis. All histograms are
normalized to a peak value of 1.  {\it Bottom:} Integrated source
counts in both subregions. The data points are the source counts
derived from the direct estimate, the black lines the results from the
$P(D)$ analysis. The grey line shows the SHADES source counts, the
grey area the source counts derived by other 870\,\micron\ studies
(see Fig.\ref{source-counts} right). }
\label{counts-region}
\end{figure}

The source counts were determined in the same manner as described in
Sect.\,\ref{sect-diffcounts} using a $P(D)$ analysis and a direct
estimate of the source counts.
%
%\begin{figure}[th]
%\centering
%\includegraphics[width=6.5cm,angle=0]{fig13.eps}
%\caption{Variation of the source counts with source surface density: {\it Top:} 
%The black and the red histograms show the pixel flux distribution for
%the dense and sparse region (Fig.\,\ref{fig-lss}).  The green curves
%show the results from our $P(D)$ analysis. All histograms are
%normalized to a peak value of 1.  {\it Bottom:} Integrated source
%counts in both subregions. The data points are the source counts
%derived from the direct estimate, the black lines the results from the
%$P(D)$ analysis. The grey line shows the SHADES source counts, the
%grey area the source counts derived by other 870\,\micron\ studies
%(see Fig.\ref{source-counts} right). }
%\label{counts-region}
%\end{figure}
%
In the top panel of Fig.\,\ref{counts-region} we show the observed
flux histograms for both regions in comparison to the fits from the
$P(D)$ analysis. For the sparse region we find that the source counts
are well described by a single power law. The best fitting model
yields a normalization at 5\,mJy half of that for the full field
($N'=47\pm 8\,{\rm deg}^{-2}\,{\rm mJy}^{-1}$) and a slope of
$\alpha=3.6\pm0.3$ -- steeper than the counts for the full field
($\alpha=3.2\pm0.2$). The $P(D)$ analysis suggests that the source counts 
of the dense region are much shallower. For a single power law we find good
matching parameters of $N'= 250\pm 20\,{\rm deg}^{-2}\,{\rm mJy}^{-1}$
and $\alpha=2.9\pm0.2$. The fitting, however, suggests that the counts
are slightly better described by a broken power law or a Schechter
function. For a broken power law the break occurs at $\approx8$\,mJy
with a normalization of $N'=100\pm 8\,{\rm deg}^{-2}\,{\rm mJy}^{-1}$.
For the slopes we find $\alpha=2.4\pm0.15$ and $\beta=4.7\pm0.6$. In
the bottom part of Fig.\,\ref{counts-region} we compare the integrated
source counts determined from the direct source count estimate and the
$P(D)$ analysis for both regions to the counts of other 870\micron\
surveys. Fig.~\ref{counts-region} shows that the source counts for the
dense region are in good agreement with results of the SHADES and
other submm surveys. This comparison demonstrates that 1)
the surface density of submm sources is not constant, but changes by a
factor $\sim3$ on angular scales of $\sim10'$ and 2) the sampling
variance is not due to a simple scaling of the number counts but
associated with changes of the shape of the source counts. The latter
finding suggests that the sampling variance is not due to (weak)
lensing by foreground mass distributions but intrinsic to the
distribution of submm sources.

\subsection{Contribution to the EBL}

 The $P(D)$ analysis of the differential source counts also provides
 an estimate on the integrated 870\,\micron\ background light that can
 be compared to the interpolated EBL at submm wavelength from {\it
 COBE FIRAS} of $\sim45$ Jy\,deg$^{-2}$
 \citep{puget96,fixsen98}. Depending on the assumed underlying source
 distribution we recover an EBL contribution of
 $29-32$\,Jy\,deg$^{-2}$ for the ECDFS (see Table\,\ref{models}). We
 note that this range is a lower limit to the true underlying
 870\,\micron\ flux as the contribution of very faint sources, which
 are expected to form an almost uniform distribution across the field
 at the spatial resolution of LABOCA, will be removed from our data in
 the correlated noise removal steps. In comparison to the values
 measured by {\it COBE} we therefore detect $>65-70$\% of the EBL for
 sources brighter than $\sim 0.5$\,mJy (which corresponds to a typical
 lower flux cutoff in our $P(D)$ analysis). These numbers demonstrate
 that it is unlikely that the integrated EBL level in the ECDFS is
 significantly lower than in other parts of the sky. This implies that
 the observed factor 2 underdensity of submm source relative to other
 deep fields is restricted to ULIRGs with far infrared luminosities of
 $>2\times10^{12}\lsol$ (assuming no lensing and $z>0.5$ for the bulk
 of our sources), while more typical star forming galaxies, which \clearpage
\noindent  dominate the EBL, are not underabundant in the field. This conclusion
 is also supported by the intensity of the EBL we find for the two
 subregions discussed in Section\,\ref{sect-varcounts}, where our
 $P(D)$ analysis does not yield a significant difference between the
 two fields.

\section{Summary and Conclusions}

We have presented a deep 870\,\micron\ survey of the ECDFS using
LABOCA on the APEX telescope at Llano de Chajnantor in Chile. This is
the largest contiguous deep submm survey to date. Our map has a highly
uniform noise level across the full $30'\times30'$ field of
1.2\,mJy$\,{\rm beam}^{-1}$ and our survey is $>95\%$ complete for
sources down to a flux limit of 6.5\,mJy. Our main findings are
summarized as follows:

\begin{itemize}

\item At the (beam smoothed) spatial resolution of $27''$ of our
survey we find that the map's noise level is affected by confusion
noise arising from faint, individually undetected SMGs. From the rms
noise as a function of integration time we derive a confusion noise of
$\sigma_{\rm c}\approx\, 0.9\,{\rm mJy}\,{\rm beam}^{-1}$.

\item We identify 126 submm sources in a search area of
1260\,arcmin$^2$ above a signal to noise threshold of $3.7\,\sigma$,
which corresponds to an expected false detection rate of 5 sources.

\item We have determined the differential and integrated source counts
using a $P(D)$ analysis and an estimate based on our source catalog.
Both results are in reasonable agreement and show that SMGs in the
ECDFS are underabundant by a factor of $\sim2$ for sources brighter
than 3\,mJy compared to the average of previous surveys. Under the
assumption that the bulk of the sources are at $z>0.5$, this implies
an underdensity of ULIRGs with $\lfir>2\times10^{12}\lsol$ compared to
other blank fields that have been observed in the submm. The source
counts are well described by a single power law with a slope of
$\alpha=3.2\pm0.2$.

\item We derive the angular two-point correlation function for the
SMGs and find clustering on angular scales $<1'$ with a significance
up to $3.4\,\sigma$. Assuming a power law dependence for the
correlation function we derive a clustering amplitude of
$A_w=0.011\pm0.0046$ or a characteristic angular scale of
$\theta_0=14''\pm7''$ for $\gamma =1.8$. Assuming a redshift
distribution similar to that observed for spectroscopically confirmed
SMGs, we derive a correlation length of $r_0=13\pm6\,h^{-1}$\,Mpc,
somewhat larger than previous estimates of the 3-D clustering of SMGs
but in agreement with the clustering derived for 24\,\micron\ selected
ULIRGs.

\item We have investigated for the first time the spatial variations
of the SMG source counts. We find that the differential source counts
in regions with an overdensity of SMGs have a different shape compared
to those with underdensities.  While the counts in underdense regions
are well fitted by a single power law with a slope of
$\alpha=3.6\pm0.3$, the counts in the overdensities are significantly
shallower with $\alpha=2.9\pm0.2$. The counts in the overdensities are
slightly better described by a broken power law or a Schechter
function. For flux densities below 8\,mJy we find a slope of
$\alpha=2.4\pm0.15$, for sources above this limit the counts are much
steeper with $\beta=4.7\pm0.6$. This may indicate an intrinsic
turn-over in the underlying luminosity function placing an upper limit
on the FIR luminosity.

\item The integrated 870\,\micron\ flux density derived from our
survey is $>29-32$\,Jy\,deg$^{-2}$ for sources brighter than
$\sim0.5$\,mJy which corresponds to $>65-70$\% of the extragalactic
background light estimated from {\it COBE} measurements. We do not
find a significant difference of the quantity between SMG over- and
underdensities. We conclude that ECDFS is underabundant of ULIRGs but
not of more typical star forming systems with lower FIR luminosities,
which dominate the extragalactic background light.  \end{itemize}

\acknowledgments 
We would like to thank the APEX staff for their aid
in carrying out the observations. APEX is operated by the
Max-Planck-Institut f\"ur Radioastronomie, the European Southern
Observatory, and the Onsala Space Observatory. IRS, KEKC and RJI
acknowledge support from STFC. JSD acknowledges the support of the
Royal Society and of STFC.

\clearpage
\begin{deluxetable}{lcccrrcc}
\tabletypesize{\small}
\tablecaption{870\micron\ LABOCA source catalog of the ECDFS.\label{sourcelist}}
\startdata
Name & RA & DEC &  \multicolumn{1}{c}{$S_{\nu\,{\rm obs}}$} &\multicolumn{1}{c}{$^{a}S_{\nu\,{\rm deboost}}$}& S/N & $^{b}$FDR & $^{c}$Remark\\ 
(IAU) & \multicolumn{2}{c}{J2000.0}& \multicolumn{1}{l}{[mJy]} &  \multicolumn{1}{l}{[mJy]}  & & &\\ 
\tableline
\tableline
LESS\,J033314.3--275611 & 03 33 14.26 & -27 56 11.2 & $ 14.7\pm1.2$ & $14.5\pm1.2$ & 12.5 & 0.0 & 1,2\\
LESS\,J033302.5--275643 & 03 33 02.50 & -27 56 43.6 & $ 12.2\pm1.2$ & $12.0\pm1.2$ & 10.3 & 0.0 & 1,2\\
LESS\,J033321.5--275520 & 03 33 21.51 & -27 55 20.2 & $ 11.9\pm1.2$ & $11.7\pm1.2$ & 10.1 & 0.0 & 1,2\\
LESS\,J033136.0--275439 & 03 31 36.01 & -27 54 39.2 & $ 11.2\pm1.2$ & $11.0\pm1.2$ & 9.7 & 0.0 & 1,2\\
LESS\,J033129.5--275907 & 03 31 29.46 & -27 59 07.3 & $ 10.1\pm1.2$ & $10.0\pm1.2$ & 8.5 & 0.0 & 1,2\\
LESS\,J033257.1--280102 & 03 32 57.14 & -28 01 02.1 & $ 9.8\pm1.2$ &  $9.7\pm1.2$ & 8.2 & 0.0 & 1,2\\
LESS\,J033315.6--274523 & 03 33 15.55 & -27 45 23.6 & $ 9.4\pm1.2$ &  $9.2\pm1.2$ & 7.9 & 0.0 & 1,2\\
LESS\,J033205.1--273108 & 03 32 05.07 & -27 31 08.8 & $ 12.1\pm1.6$ & $11.7\pm1.6$ & 7.8 & 0.0 & 1,2\\
LESS\,J033211.3--275210 & 03 32 11.29 & -27 52 10.4 & $ 9.4\pm1.2$ &  $9.2\pm1.2$ & 7.7 & 0.0 & 1,2\\
LESS\,J033219.0--275219 & 03 32 19.02 & -27 52 19.4 & $ 9.3\pm1.2$ &  $9.1\pm1.2$ & 7.6 & 0.0 & 1,2\\
LESS\,J033213.6--275602 & 03 32 13.58 & -27 56 02.5 & $ 9.2\pm1.2$ &  $9.1\pm1.2$ & 7.6 & 0.0 & 1,2\\
LESS\,J033248.1--275414 & 03 32 48.12 & -27 54 14.7 & $ 8.9\pm1.2$ &  $8.8\pm1.2$ & 7.2 & 0.0 & 1,2\\
LESS\,J033249.2--274246 & 03 32 49.23 & -27 42 46.6 & $ 8.9\pm1.2$ &  $8.8\pm1.2$ & 7.2 & 0.0 & 1,2\\
LESS\,J033152.6--280320 & 03 31 52.64 & -28 03 20.4 & $ 9.5\pm1.3$ &  $9.3\pm1.3$ & 7.2 & 0.0 & 1,2\\
LESS\,J033333.4--275930 & 03 33 33.36 & -27 59 30.1 & $ 9.1\pm1.3$ &  $8.9\pm1.3$ & 7.0 & 0.0 & 1,2\\
LESS\,J033218.9--273738 & 03 32 18.89 & -27 37 38.7 & $ 8.2\pm1.2$ &  $8.1\pm1.2$ & 6.9 & 0.0 & 1,2\\
LESS\,J033207.6--275123 & 03 32 07.59 & -27 51 23.0 & $ 7.8\pm1.2$ &  $7.6\pm1.3$ & 6.4 & 0.0 & 1\\
LESS\,J033205.1--274652 & 03 32 05.12 & -27 46 52.1 & $ 7.7\pm1.2$ &  $7.5\pm1.2$ & 6.3 & 0.0 & 1,2\\
LESS\,J033208.1--275818 & 03 32 08.10 & -27 58 18.7 & $ 7.5\pm1.2$ &  $7.3\pm1.2$ & 6.2 & 0.0 & 2\\
LESS\,J033316.6--280018 & 03 33 16.56 & -28 00 18.8 & $ 7.5\pm1.2$ &  $7.2\pm1.2$ & 6.2 & 0.0 & 1\\
LESS\,J033329.9--273441 & 03 33 29.93 & -27 34 41.7 & $ 7.9\pm1.3$ &  $7.6\pm1.3$ & 6.1 & 0.0 & 2\\
LESS\,J033147.0--273243 & 03 31 47.02 & -27 32 43.0 & $ 8.5\pm1.4$ &  $8.0\pm1.5$ & 5.9 & 0.0 & 1,2\\
LESS\,J033212.1--280508 & 03 32 12.11 & -28 05 08.5 & $ 8.8\pm1.5$ &  $8.2\pm1.5$ & 5.9 & 0.0 & 1,2\\
LESS\,J033336.8--274401 & 03 33 36.79 & -27 44 01.0 & $ 7.8\pm1.3$ &  $7.4\pm1.4$ & 5.9 & 0.0 & 2\\
LESS\,J033157.1--275940 & 03 31 57.05 & -27 59 40.8 & $ 7.0\pm1.2$ &  $6.7\pm1.3$ & 5.8 & 0.0 & 1\\
LESS\,J033136.9--275456 & 03 31 36.90 & -27 54 56.1 & $ 6.8\pm1.2$ &  $6.6\pm1.2$ & 5.8 & 0.0 & 1\\
LESS\,J033149.7--273432 & 03 31 49.73 & -27 34 32.7 & $ 7.6\pm1.3$ &  $7.2\pm1.4$ & 5.8 & 0.0 & 1\\
LESS\,J033302.9--274432 & 03 33 02.92 & -27 44 32.6 & $ 7.0\pm1.3$ &  $6.7\pm1.3$ & 5.6 & 0.0 & 1,2\\
LESS\,J033336.9--275813 & 03 33 36.90 & -27 58 13.0 & $ 7.6\pm1.4$ &  $7.1\pm1.4$ & 5.6 & 0.0 & 1\\
LESS\,J033344.4--280346 & 03 33 44.37 & -28 03 46.1 & $ 9.7\pm1.7$ &  $8.7\pm1.8$ & 5.6 & 0.0 & 2\\
LESS\,J033150.0--275743 & 03 31 49.96 & -27 57 43.9 & $ 6.7\pm1.2$ &  $6.3\pm1.3$ & 5.5 & 0.0 & 1,2\\
LESS\,J033243.6--274644 & 03 32 43.57 & -27 46 44.0 & $ 6.8\pm1.2$ &  $6.4\pm1.3$ & 5.5 & 0.0 & 1\\
LESS\,J033149.8--275332 & 03 31 49.78 & -27 53 32.9 & $ 6.8\pm1.3$ &  $6.4\pm1.3$ & 5.5 & 0.0 & 0\\
LESS\,J033217.6--275230 & 03 32 17.64 & -27 52 30.3 & $ 6.8\pm1.3$ &  $6.3\pm1.3$ & 5.4 & 0.0 & 0\\
LESS\,J033110.4--273714 & 03 31 10.35 & -27 37 14.8 & $ 9.1\pm1.7$ &  $8.1\pm1.8$ & 5.4 & 0.0 & 2\\
LESS\,J033149.2--280208 & 03 31 49.15 & -28 02 08.7 & $ 6.9\pm1.3$ &  $6.4\pm1.4$ & 5.4 & 0.0 & 2\\
LESS\,J033336.0--275347 & 03 33 36.04 & -27 53 47.6 & $ 7.3\pm1.4$ &  $6.7\pm1.5$ & 5.3 & 0.0 & 0\\
LESS\,J033310.2--275641 & 03 33 10.20 & -27 56 41.5 & $ 6.4\pm1.2$ &  $6.0\pm1.3$ & 5.2 & 0.0 & 0\\
LESS\,J033144.9--273435 & 03 31 44.90 & -27 34 35.4 & $ 6.8\pm1.3$ &  $6.2\pm1.4$ & 5.2 & 0.0 & 0\\
LESS\,J033246.7--275120 & 03 32 46.74 & -27 51 20.9 & $ 6.4\pm1.2$ &  $5.9\pm1.3$ & 5.2 & 0.0 & 2\\
LESS\,J033110.5--275233 & 03 31 10.47 & -27 52 33.2 & $ 8.7\pm1.7$ &  $7.6\pm1.9$ & 5.2 & 0.0 & 1\\
LESS\,J033231.0--275858 & 03 32 31.02 & -27 58 58.1 & $ 6.4\pm1.2$ &  $5.8\pm1.4$ & 5.1 & 0.0 & 2\\
LESS\,J033307.0--274801 & 03 33 07.00 & -27 48 01.0 & $ 6.4\pm1.3$ &  $5.9\pm1.4$ & 5.1 & 0.0 & 0\\
LESS\,J033131.0--273238 & 03 31 30.96 & -27 32 38.5 & $ 7.5\pm1.5$ &  $6.7\pm1.6$ & 5.1 & 0.0 & 0\\
LESS\,J033225.7--275228 & 03 32 25.71 & -27 52 28.5 & $ 6.3\pm1.2$ &  $5.8\pm1.4$ & 5.1 & 0.0 & 1\\
LESS\,J033336.8--273247 & 03 33 36.80 & -27 32 47.0 & $ 8.2\pm1.6$ &  $7.2\pm1.8$ & 5.1 & 0.0 & 0\\
LESS\,J033256.0--273317 & 03 32 56.00 & -27 33 17.7 & $ 7.0\pm1.4$ &  $6.3\pm1.5$ & 5.1 & 0.0 & 0\\
LESS\,J033237.8--273202 & 03 32 37.77 & -27 32 02.0 & $ 7.7\pm1.5$ &  $6.8\pm1.7$ & 5.1 & 0.0 & 0\\
LESS\,J033124.5--275040 & 03 31 24.45 & -27 50 40.9 & $ 6.6\pm1.3$ &  $5.9\pm1.4$ & 5.1 & 0.0 & 1,2\\
LESS\,J033141.2--274441 & 03 31 41.15 & -27 44 41.5 & $ 6.1\pm1.2$ &  $5.6\pm1.3$ & 5.0 & 0.0 & 1,2\\
LESS\,J033144.8--274425 & 03 31 44.81 & -27 44 25.1 & $ 6.2\pm1.2$ &  $5.6\pm1.3$ & 5.0 & 0.0 & 2\\
LESS\,J033128.5--275601 & 03 31 28.51 & -27 56 01.3 & $ 6.2\pm1.3$ &  $5.6\pm1.4$ & 4.9 & 0.0 & 1\\
LESS\,J033159.1--275435 & 03 31 59.12 & -27 54 35.5 & $ 6.2\pm1.3$ &  $5.6\pm1.4$ & 4.9 & 0.0 & 2\\
LESS\,J033243.6--273353 & 03 32 43.61 & -27 33 53.6 & $ 6.8\pm1.4$ &  $6.0\pm1.5$ & 4.9 & 0.0 & 2\\
LESS\,J033302.2--274033 & 03 33 02.20 & -27 40 33.6 & $ 6.1\pm1.2$ &  $5.5\pm1.4$ & 4.9 & 0.0 & 0\\
LESS\,J033153.2--273936 & 03 31 53.17 & -27 39 36.1 & $ 6.0\pm1.2$ &  $5.4\pm1.4$ & 4.9 & 0.0 & 1\\
LESS\,J033152.0--275329 & 03 31 51.97 & -27 53 29.7 & $ 6.1\pm1.3$ &  $5.5\pm1.4$ & 4.9 & 0.0 & 0\\
LESS\,J033225.8--273306 & 03 32 25.79 & -27 33 06.7 & $ 6.7\pm1.4$ &  $5.9\pm1.6$ & 4.8 & 0.0 & 0\\
LESS\,J033303.9--274412 & 03 33 03.87 & -27 44 12.2 & $ 6.0\pm1.3$ &  $5.3\pm1.4$ & 4.8 & 0.0 & 0\\
LESS\,J033317.5--275121 & 03 33 17.47 & -27 51 21.5 & $ 5.8\pm1.2$ &  $5.2\pm1.4$ & 4.8 & 0.1 & 1\\
LESS\,J033245.6--280025 & 03 32 45.63 & -28 00 25.3 & $ 5.9\pm1.2$ &  $5.2\pm1.4$ & 4.7 & 0.1 & 0\\
LESS\,J033236.4--273452 & 03 32 36.41 & -27 34 52.5 & $ 6.1\pm1.3$ &  $5.4\pm1.5$ & 4.7 & 0.1 & 2\\
LESS\,J033308.5--280044 & 03 33 08.46 & -28 00 44.3 & $ 6.0\pm1.3$ &  $5.3\pm1.4$ & 4.7 & 0.1 & 2\\
LESS\,J033201.0--280025 & 03 32 01.00 & -28 00 25.6 & $ 5.8\pm1.2$ &  $5.1\pm1.4$ & 4.7 & 0.1 & 1\\
LESS\,J033252.4--273527 & 03 32 52.40 & -27 35 27.7 & $ 5.9\pm1.3$ &  $5.2\pm1.4$ & 4.7 & 0.1 & 0\\
LESS\,J033331.7--275406 & 03 33 31.69 & -27 54 06.1 & $ 6.1\pm1.3$ &  $5.3\pm1.5$ & 4.7 & 0.1 & 0\\
LESS\,J033243.3--275517 & 03 32 43.28 & -27 55 17.9 & $ 5.9\pm1.3$ &  $5.2\pm1.4$ & 4.7 & 0.1 & 0\\
LESS\,J033233.4--273918 & 03 32 33.44 & -27 39 18.5 & $ 5.8\pm1.3$ &  $5.1\pm1.4$ & 4.7 & 0.1 & 1\\
LESS\,J033134.3--275934 & 03 31 34.26 & -27 59 34.3 & $ 5.7\pm1.2$ &  $5.0\pm1.3$ & 4.7 & 0.1 & 1\\
LESS\,J033144.0--273832 & 03 31 43.97 & -27 38 32.5 & $ 5.7\pm1.2$ &  $5.0\pm1.4$ & 4.6 & 0.1 & 2\\
LESS\,J033306.3--273327 & 03 33 06.29 & -27 33 27.7 & $ 6.6\pm1.4$ &  $5.6\pm1.6$ & 4.6 & 0.1 & 0\\
LESS\,J033240.4--273802 & 03 32 40.40 & -27 38 02.5 & $ 5.7\pm1.2$ &  $5.0\pm1.4$ & 4.6 & 0.1 & 1\\
LESS\,J033229.3--275619 & 03 32 29.33 & -27 56 19.3 & $ 5.8\pm1.3$ &  $5.1\pm1.4$ & 4.6 & 0.1 & 0\\
LESS\,J033309.3--274809 & 03 33 09.34 & -27 48 09.9 & $ 5.8\pm1.3$ &  $5.1\pm1.4$ & 4.6 & 0.1 & 0\\
LESS\,J033126.8--275554 & 03 31 26.83 & -27 55 54.6 & $ 5.8\pm1.3$ &  $5.1\pm1.4$ & 4.6 & 0.1 & 2\\
LESS\,J033332.7--275957 & 03 33 32.67 & -27 59 57.2 & $ 6.0\pm1.3$ &  $5.1\pm1.5$ & 4.5 & 0.1 & 0\\
LESS\,J033157.2--275633 & 03 31 57.23 & -27 56 33.2 & $ 5.6\pm1.3$ &  $4.8\pm1.4$ & 4.4 & 0.2 & 0\\
LESS\,J033340.3--273956 & 03 33 40.30 & -27 39 56.9 & $ 6.2\pm1.4$ &  $5.1\pm1.7$ & 4.4 & 0.3 & 1\\
LESS\,J033221.3--275623 & 03 32 21.25 & -27 56 23.5 & $ 5.5\pm1.3$ &  $4.7\pm1.4$ & 4.4 & 0.3 & 2\\
LESS\,J033142.2--274834 & 03 31 42.23 & -27 48 34.4 & $ 5.4\pm1.2$ &  $4.6\pm1.4$ & 4.4 & 0.3 & 2\\
LESS\,J033127.5--274440 & 03 31 27.45 & -27 44 40.4 & $ 5.7\pm1.3$ &  $4.8\pm1.5$ & 4.4 & 0.3 & 0\\
LESS\,J033253.8--273810 & 03 32 53.77 & -27 38 10.9 & $ 5.4\pm1.2$ &  $4.5\pm1.4$ & 4.4 & 0.3 & 0\\
LESS\,J033308.9--280522 & 03 33 08.92 & -28 05 22.0 & $ 6.7\pm1.5$ &  $5.3\pm1.8$ & 4.4 & 0.3 & 0\\
LESS\,J033154.2--275109 & 03 31 54.22 & -27 51 09.8 & $ 5.5\pm1.3$ &  $4.6\pm1.4$ & 4.3 & 0.4 & 0\\
LESS\,J033110.3--274503 & 03 31 10.28 & -27 45 03.1 & $ 8.2\pm1.6$ &  $6.0\pm2.4$ & 4.3 & 0.4 & 1\\
LESS\,J033114.9--274844 & 03 31 14.90 & -27 48 44.3 & $ 6.5\pm1.5$ &  $5.1\pm1.8$ & 4.3 & 0.4 & 0\\
LESS\,J033251.1--273143 & 03 32 51.09 & -27 31 43.0 & $ 6.7\pm1.6$ &  $5.3\pm1.9$ & 4.3 & 0.5 & 0\\
LESS\,J033155.2--275345 & 03 31 55.19 & -27 53 45.3 & $ 5.4\pm1.3$ &  $4.5\pm1.4$ & 4.3 & 0.5 & 0\\
LESS\,J033248.4--280023 & 03 32 48.44 & -28 00 23.8 & $ 5.3\pm1.2$ &  $4.4\pm1.4$ & 4.3 & 0.5 & 0\\
LESS\,J033243.7--273554 & 03 32 43.65 & -27 35 54.1 & $ 5.4\pm1.3$ &  $4.5\pm1.5$ & 4.2 & 0.6 & 0\\
LESS\,J033135.3--274033 & 03 31 35.25 & -27 40 33.7 & $ 5.3\pm1.3$ &  $4.4\pm1.4$ & 4.2 & 0.6 & 0\\
LESS\,J033138.4--274336 & 03 31 38.36 & -27 43 36.0 & $ 5.2\pm1.2$ &  $4.3\pm1.4$ & 4.2 & 0.6 & 0\\
LESS\,J033110.8--275607 & 03 31 10.84 & -27 56 07.2 & $ 6.9\pm1.7$ &  $5.2\pm2.0$ & 4.2 & 0.6 & 0\\
LESS\,J033307.3--275805 & 03 33 07.27 & -27 58 05.0 & $ 5.3\pm1.3$ &  $4.4\pm1.4$ & 4.2 & 0.7 & 0\\
LESS\,J033241.7--275846 & 03 32 41.74 & -27 58 46.1 & $ 5.2\pm1.3$ &  $4.3\pm1.4$ & 4.2 & 0.7 & 0\\
LESS\,J033313.0--275556 & 03 33 13.03 & -27 55 56.8 & $ 5.2\pm1.2$ &  $4.3\pm1.4$ & 4.2 & 0.7 & 0\\
LESS\,J033313.7--273803 & 03 33 13.65 & -27 38 03.4 & $ 5.1\pm1.2$ &  $4.2\pm1.4$ & 4.2 & 0.8 & 0\\
LESS\,J033130.2--275726 & 03 31 30.22 & -27 57 26.0 & $ 5.1\pm1.3$ &  $4.2\pm1.4$ & 4.1 & 1.0 & 0\\
LESS\,J033251.5--275536 & 03 32 51.45 & -27 55 36.0 & $ 5.3\pm1.3$ &  $4.3\pm1.4$ & 4.1 & 1.0 & 0\\
LESS\,J033111.3--280006 & 03 31 11.32 & -28 00 06.2 & $ 6.4\pm1.6$ &  $4.8\pm1.9$ & 4.1 & 1.0 & 0\\
LESS\,J033151.5--274552 & 03 31 51.47 & -27 45 52.1 & $ 5.1\pm1.3$ &  $4.2\pm1.4$ & 4.1 & 1.1 & 1\\
LESS\,J033335.6--274020 & 03 33 35.61 & -27 40 20.1 & $ 5.4\pm1.3$ &  $4.3\pm1.5$ & 4.1 & 1.2 & 0\\
LESS\,J033325.4--273400 & 03 33 25.35 & -27 34 00.4 & $ 5.5\pm1.3$ &  $4.3\pm1.5$ & 4.1 & 1.2 & 0\\
LESS\,J033258.5--273803 & 03 32 58.46 & -27 38 03.0 & $ 4.9\pm1.2$ &  $4.0\pm1.4$ & 4.1 & 1.3 & 0\\
LESS\,J033115.8--275313 & 03 31 15.78 & -27 53 13.1 & $ 6.0\pm1.5$ &  $4.6\pm1.7$ & 4.1 & 1.3 & 0\\
LESS\,J033140.1--275631 & 03 31 40.09 & -27 56 31.4 & $ 4.9\pm1.2$ &  $4.0\pm1.4$ & 4.0 & 1.4 & 2\\
LESS\,J033130.9--275150 & 03 31 30.85 & -27 51 50.9 & $ 5.0\pm1.3$ &  $4.0\pm1.4$ & 4.0 & 1.4 & 0\\
LESS\,J033316.4--275033 & 03 33 16.42 & -27 50 33.1 & $ 5.0\pm1.2$ &  $4.0\pm1.4$ & 4.0 & 1.5 & 0\\
LESS\,J033328.1--274157 & 03 33 28.08 & -27 41 57.0 & $ 5.0\pm1.3$ &  $4.0\pm1.4$ & 4.0 & 1.6 & 0\\
LESS\,J033122.6--275417 & 03 31 22.64 & -27 54 17.2 & $ 5.3\pm1.3$ &  $4.1\pm1.5$ & 4.0 & 1.8 & 0\\
LESS\,J033325.6--273423 & 03 33 25.58 & -27 34 23.0 & $ 5.2\pm1.3$ &  $4.1\pm1.5$ & 4.0 & 1.9 & 0\\
LESS\,J033249.3--273112 & 03 32 49.28 & -27 31 12.3 & $ 6.5\pm1.7$ &  $4.6\pm2.0$ & 4.0 & 2.0 & 0\\
LESS\,J033236.4--275845 & 03 32 36.42 & -27 58 45.9 & $ 5.0\pm1.3$ &  $3.9\pm1.4$ & 3.9 & 2.0 & 0\\
LESS\,J033150.8--274438 & 03 31 50.81 & -27 44 38.5 & $ 4.9\pm1.3$ &  $3.9\pm1.4$ & 3.9 & 2.4 & 0\\
LESS\,J033349.7--274239 & 03 33 49.71 & -27 42 39.2 & $ 7.4\pm1.6$ &  $4.6\pm2.4$ & 3.9 & 2.9 & 0\\
LESS\,J033154.4--274525 & 03 31 54.42 & -27 45 25.5 & $ 4.9\pm1.3$ &  $3.8\pm1.4$ & 3.8 & 3.0 & 0\\
LESS\,J033128.0--273925 & 03 31 28.02 & -27 39 25.2 & $ 5.0\pm1.3$ &  $3.8\pm1.4$ & 3.8 & 3.1 & 0\\
LESS\,J033121.8--274936 & 03 31 21.81 & -27 49 36.8 & $ 5.2\pm1.4$ &  $3.8\pm1.5$ & 3.8 & 3.7 & 0\\
LESS\,J033256.5--280319 & 03 32 56.51 & -28 03 19.1 & $ 5.1\pm1.4$ &  $3.8\pm1.5$ & 3.8 & 3.8 & 0\\
LESS\,J033328.5--275655 & 03 33 28.45 & -27 56 55.9 & $ 4.9\pm1.3$ &  $3.7\pm1.5$ & 3.8 & 3.8 & 0\\
LESS\,J033333.3--273449 & 03 33 33.32 & -27 34 49.3 & $ 5.2\pm1.4$ &  $3.8\pm1.6$ & 3.8 & 3.9 & 0\\
LESS\,J033139.6--274120 & 03 31 39.62 & -27 41 20.4 & $ 4.7\pm1.2$ &  $3.6\pm1.5$ & 3.8 & 4.0 & 0\\
LESS\,J033330.9--275349 & 03 33 30.88 & -27 53 49.3 & $ 4.9\pm1.3$ &  $3.7\pm1.6$ & 3.8 & 4.2 & 0\\
LESS\,J033203.6--273605 & 03 32 03.59 & -27 36 05.0 & $ 4.6\pm1.2$ &  $3.5\pm1.4$ & 3.7 & 4.7 & 0\\
LESS\,J033146.0--274621 & 03 31 46.02 & -27 46 21.2 & $ 4.7\pm1.3$ &  $3.6\pm1.4$ & 3.7 & 4.7 & 0\\
LESS\,J033209.8--274102 & 03 32 09.76 & -27 41 02.0 & $ 4.7\pm1.3$ &  $3.6\pm1.4$ & 3.7 & 4.9 & 0\\
\enddata
\tablenotetext{a}{Deboosted fluxes depend on the source count model (see Sect.\,\ref{deboost}) and are only correct in a statistical sense.}
\tablenotetext{b}{Expected number of false detections for all sources including the corresponding entry in the Table.}
\tablenotetext{c}{The ''Remark'' entry indicates if a source is detected in the two submaps calculated by splitting
the data into half (see \ref{sourcetests}). 1,2: the source is detection in both submaps, 1 (2): the 
source is detected in submap 1 (2), 0 the source is not detected in any submap.}
\end{deluxetable}


\clearpage


\begin{thebibliography}{}
\bibitem[Austermann \etal(2009)]{austermann09}Austermann, J.E., \etal\ 2009, MNRAS,393, 1573
\bibitem[Barger \etal(1999)]{barger99}Barger, A.J., Cowie, L.L., Sanders, D.B., 1999, ApJ, 518, 5
\bibitem[Beckwith \etal(2006)]{beckwith06}Beckwith, S.V.W, \etal\  2006, AJ, 132, 1729
\bibitem[Beelen \etal(2008)]{beelen08}Beelen, A., \etal\ 2008, A\&A., 485, 645
\bibitem[Bertoldi \etal(2007)]{bertoldi07}Bertoldi, F., \etal\ 2007, ApJS, 172, 132
\bibitem[Blain \etal(1999)]{blain99}Blain, A.W., Smail, I., Ivison, R.J., \& Kneib, J.P. 1999, MNRAS, 302, 632
\bibitem[Blain \etal(2004)]{blain04}Blain, A.W., Chapman, S.C., Smail, I., Ivison, R. 2004, ApJ, 611, 725
\bibitem[Blanc \etal(2008)]{blanc08}Blanc, G,A., \etal\ 2008, ApJ, 681, 1099
\bibitem[Borys \etal(2002)]{borys02} Borys, C., Chapman, S.C., Halpern, M., Scott, D., 2002, MNRAS, 330, 63
\bibitem[Borys \etal(2003)]{borys03} Borys, C., Chapman, S.C., Halpern, M., Scott, D., 2003, MNRAS, 334, 385
\bibitem[Brainerd \& Smail (1998)]{brainerd98}Brainerd, T.,G., Smail, I. 1998, ApJ, 494, 137
\bibitem[Caldwell \etal(2008)]{caldwell08}Caldwell, J.A.R., \etal\ 2008, ApJS, 174, 136
\bibitem[Chapman \etal(2001)]{chapman01}Chapman, S.C., Lewis, G.F., Scott, D., Richards, E., Borys, C., Steidel, C.C., Adelberger, K.L., \& Shapley, A.E.2001, ApJ, 548, 17
\bibitem[Chapman \etal(2005)]{chapman05}Chapman, S.C., Blain, A.W., Smail, I., \& Ivison, R.J. 2005, ApJ, 622, 772
\bibitem[Condon (1974)]{condon74}Condon, J.J. 1974, ApJ, 188, 279
\bibitem[Coppin \etal(2005)]{coppin05} Coppin, K., Halper, M., Scott, D., Borys, C.,\& Chapman, S., 2005, MNRAS 357, 1022
\bibitem[Coppin \etal(2006)]{coppin06} Coppin, K., \etal\ 2006, MNRAS 372, 1621
\bibitem[Coppin \etal(2009)]{coppin09} Coppin, K., \etal\ 2009, MNRAS 395, 1905
\bibitem[Cowie \etal(2002)]{cowie02} Cowie, L.L, Barger, A.J., Kneib, J.P., 2002, AJ, 123, 2197
\bibitem[Daddi \etal(2000)]{daddi00}Daddi, E., Cimatti, A., Pozzetti, L., Hoekstra, H., R\"ottgering, H.J.A., Renzini, A., Zamorani, G., \& Mannucci, F . 2000, A\&A, 361, 535
\bibitem[Damen \etal(2009)]{damen09}Damen, M., Labbe, I., Franx, M., van Dokkum, P.G., Taylor, E.N., \& Gawiser, E.J. 2009,  ApJ, 690, 937
\bibitem[Devlin \etal(2009)]{devlin09}Devlin, M.J., \etal\ 2009, Nature, 458, 737 
\bibitem[Dwelly \& Page(2006)]{dwelly06} Dwelly, T., \& Page, M.J. 2006, MNRAS, 372, 1755	
\bibitem[Farrah \etal(2006)]{farrah06}Farrah, D., \etal\ 2006, ApJ, 641, 17
\bibitem[Fixsen \etal(1998)]{fixsen98}Fixsen, D.J., Dwek, E., Mather, J.C., Bennett, C.L. \& Shafer, R.A. 1998, ApJ, 508, 123
\bibitem[Gawiser \etal(2006)]{gawiser06}Gawiser, E., \etal\  2006, ApJS, 162, 1
\bibitem[Giacconi \etal(2002)]{giacconi02}Giacconi, R., \etal\ 2002, ApJS, 139, 369
\bibitem[Giavalisco \etal(2004)]{giovalisco04}Giavalisco, M., \etal\ 2004, ApJ, 600, 93
\bibitem[G\"usten \etal(2006)]{guesten06} G\"usten, R., Nyman, L.A., Schilke, P., Menten, K.M., Cesarsky C., \& Booth R. 2006, A\&A, 454, 13
\bibitem[Gutterman \etal (2006)]{gutterman06}Gutterman, Z., Pinkas, B., \&  Reinman, T. 2003, IEEE Symposium on Security and Privacy, 38, 371
\bibitem[Greve \etal(2004)]{greve04} Greve, T.R., Ivison, R.J., Bertoldi, F., \etal\ 2004, MNRAS, 354, 779
\bibitem[Greve \etal(2008)]{greve08}Greve, T.R., Pope, A., Scott, D., Ivison, R.J., Borys, C., Conselice, C.J., \&  Bertoldi, F. 2008, MNRAS, 2008, 389, 1489
\bibitem[Greve \etal(2009)]{greve09}Greve, T.R., \etal\ 2009, submitted to ApJ (astro-ph/0904.0028)
\bibitem[Hartley \etal(2008)]{hartley08}Hartley, W.G., \etal\ 2008, MNRAS, 391, 1301
\bibitem[Hogg \& Turner(1998)]{hogg98}Hogg, D.W., \& Turner, E.L. 1998, PASP, 110, 727
\bibitem[Hopkins \etal(2002)]{hopkins02}Hopkins, A.M., Miller, C.J., Connolly, A.J., Genovese, C., Nichol, R.C., \&  Wasserman, L. 2002, AJ, 123, 1086
\bibitem[Hughes \etal(1998)]{hughes98}Hughes, D., \etal\ 1998, Natur, 394, 241
\bibitem[Isaak \etal(2002)]{isaak02}Isaak, K.,G., Priddey, R.,S., McMahon, R.,G., Omont, A., Peroux, C., Sharp, R.,G. \& Withington, S. 2002, MNRAS, 329, 149
\bibitem[Ivison \etal(1998)]{ivison98} Ivison, R.J., Smail, I., Le Borgne, J.-F., Blain, A. W., Kneib, J.-P., Bezecourt, J., Kerr, T. H., \& Davies, J. K. 1998, MNRAS, 298, 583
\bibitem[Ivison \etal(2000)]{ivison00}Ivison, R.J., Dunlop, J.S., Smail, I., Dey, A., Liu, M.C., \& Graham, J.R. 2000, ApJ, 542, 27
\bibitem[Ivison \etal(2007)]{ivison07}Ivison, R.J., \etal\ 2007, MNRAS, 380, 199
\bibitem[Ivison \etal(2009)]{ivison09}Ivison, R.J., \etal\ 2009, submitted to MNRAS (astro-ph/0910.1091)
\bibitem[Kneib \etal(2004)]{kneib04}Kneib, J., van der Werf, P.P., Kraiberg Knudsen, K., Smail, I., Blain, A., 
Frayer, D., Barnard, V., \& Ivison, R. 2004, MNRAS, 349, 1211 
\bibitem[Knudsen \etal(2008)]{knudsen08}Knudsen, K.K., van der Werf, P.P., \&  Kneib, J.P. 2008, MNRAS, 384, 1611
\bibitem[Kov\'{a}cs \etal(2006)]{kovacs06}Kov\'{a}cs, A., Chapman, S.C., Dowell, C.D., Blain, A.W., Ivison, R.J., Smail, I., \& Phillips, T.G. 2006, ApJ, 650, 592
\bibitem[Kov\'{a}cs(2008)]{crush} Kov\'{a}cs, A.\ 2008, Proc.\ SPIE, 7020, Millimeter and Submillimeter Detectors for Astronomy, ed.~W.D.~Duncan, W.S.~Holland, S.~Withington, \& J.~Zmuidzinas, 45
\bibitem[Landy \& Szalay(1993)]{landy93}Landy, S.D.\& Szalay, A.S. 1993, ApJ, 412, 64
\bibitem[Laurent \etal(2005)]{laurent05}Laurent, G.T., \etal\ 2005, ApJ, 623 742
\bibitem[Le Floch \etal(2005)]{lefloch05}Le Floch, E., \etal\  2005, ApJ, 632, 169
\bibitem[Lehmer \etal(2005)]{lehmer05}Lehmer, B.D., \etal\  2005, ApJS, 161, 21
\bibitem[Luo \etal(2008)]{luo08}Luo, B., \etal\ 2008, ApJS, 179, 19
\bibitem[Lutz \etal(2009)]{lutz09}Lutz, D., \etal\ 2009, in prep.
\bibitem[Maloney \etal(2005)]{maloney05}Maloney, P.R., \etal\ 2005, ApJ, 635, 1044
\bibitem[Marchesini \etal(2007)]{marchesini07}Marchesini, D., van Dokkum, P., Quadri, R., Rudnick, G., Franx, M., Lira, P., Wuyts, S., Gawiser, E., Christlein, D., \&  Toft, S.  2007, ApJ, 656, 42
\bibitem[Matarrese \etal(1997)]{matarrese97}Matarrese, S., Coles, P., Lucchin, F., \&  Moscardini, L 1997, MNRAS, 286, 115
\bibitem[Miller \etal(2008)]{miller08}Miller, N.A., Fomalont, E.B., Kellermann, K.I., Mainieri, V., Norman, C., Padovani, P., Rosati, P.,\& Tozzi, P. 2008, ApJS, 179, 114
\bibitem[Overzier \etal(2003)]{overzier03}Overzier, R.A., R\"ottgering, H.J.A., Rengelink, R.B., \& Wilman R.J. 2003 A\&A, 405, 53
\bibitem[Peacock \etal(2000)]{peacock00}Peacock, J. A., \etal\ 2000, MNRAS, 318, 535
\bibitem[Perera \etal(2008)]{perera08}Perera, T.A., Chapin, E.L., Austermann, J.E., \etal\, 2008, MNRAS, 391, 1227
\bibitem[Pope \etal(2006)]{pope06}Pope, A., \etal\ 2006, MNRAS, 370, 1185
\bibitem[Press \etal(1986)]{press86} Press W.\ H.\, Flannery B.\ P.\, \& Teukolsky, S.\ A.\ 1986, Numerical recipes. The art of scientific computing (Cambridge: University Press)
\bibitem[Puget \etal(1996)]{puget96}Puget, J.L., Abergel, A., Bernard, J.P., Boulanger, F., Burton, W.B., Desert, F.X.,\& 
Hartmann, D. 1996,  A\&A, 308, 5
\bibitem[Quadri \etal(2008)]{quadri08}Quadri, R.,F., Williams, R.J., Lee, K.S.; Franx, M., van Dokkum, P., \& Brammer, G.B. 2008, ApJ, 685, 1
\bibitem[Sanders \& Mirabel (1996)]{sanders96}Sanders, D.B. \& Mirabel, I.F. 1996, ARA\&A, 34, 749
\bibitem[Scott \etal(2002)]{scott02}Scott, S.E., \etal\ 2002, MNRAS, 331, 817 
\bibitem[Scott \etal(2006)]{scott06}Scott, S.E., Dunlop, J.S., \& Serjeant, S. 2006, MNRAS, 370, 1057
\bibitem[Scott \etal(2008)]{scott08}Scott, K.S., \etal\ 2008, MNRAS, 385, 2225
\bibitem[Serjeant \etal(2003)]{serjeant03}Serjeant, S., \etal\ 2003, MNRAS, 344, 887
\bibitem[Siringo \etal(2009)]{siringo09} Siringo, G., \etal\ 2009, A\&A, 497, 945
\bibitem[Smail, Ivison \& Blain(1997)]{smail97}Smail, I., Ivison, R.J, \& Blain, A.W., 1997, ApJ, 490, 5
\bibitem[Smail \etal(2002)]{smail02}Smail, I., Ivison, R.J., Blain, A.W., \& Kneib, J.P. 2002, MNRAS, 331, 495
\bibitem[Stevens, \etal(2003)]{stevens03}Stevens, J.A., \etal\ 2003, Natur, 425, 264
\bibitem[Stevens, \etal(2003, 2004)]{stevens04}Stevens, J.A., Page, M.J., Ivison, R.J., Smail, I. \&  Carrera, F.J. 2004, ApJ, 604, 17
\bibitem[Swinbank \etal(2008)]{swinbank08}Swinbank, A.M., \etal\  2008, MNRAS, 391, 420
\bibitem[Tacconi \etal(2008)]{tacconi08} Tacconi, L.J., \etal\ 2008 ApJ, 680, 246
\bibitem[Taylor \etal(2009)]{taylor09}Talor, E.N., \etal\ 2009, ApJSS in press (astro-ph/0903.3051)
\bibitem[van Dokkum \etal(2006)]{dokkum06}van Dokkum, P.G., \etal\  2006 ApJ, 638 59
\bibitem[van Kampen \etal(2005)]{vankampen05}van Kampen E. \etal\ 2005, MNRAS, 359, 469 
\bibitem[Webb \etal(2003)]{webb03}Webb, T.M., Eales, S.A., Lilly, S.J., Clements, D.L., Dunne, L., Gear, W.K., Ivison, R.J., Flores, H., \& Yun, M. 2003, ApJ, 587, 41
\bibitem[Wolf \etal(2004)]{wolf04}Wolf, C., \etal\  2004, A\&A, 421, 913
\bibitem[Wolf \etal(2008)]{wolf08}Wolf, C., Hildebrandt, H., Taylor, E.N., \& Meisenheimer, K. 2008, A\&A, 492, 933
\bibitem[Younger \etal(2008)]{younger08}Younger, J.D., \etal\ 2008, ApJ, 688, 59
\end{thebibliography}
\end{document}